\documentclass[prd,preprint,superscriptaddress,preprintnumbers,eqsecnum,showpacs,nofootinbib,nobibnotes]{revtex4-1}
\usepackage{amsfonts,amsmath,mathbbol,amssymb,bm,natbib}

\usepackage{amsmath}
\usepackage{amssymb}
\usepackage{graphicx}
\usepackage{color}
\usepackage{xspace}
\usepackage{array,makecell}

\usepackage[mathscr,scaled=1.15]{urwchancal}
\DeclareFontFamily{OT1}{pzc}{}
\DeclareFontShape{OT1}{pzc}{m}{it}%
{<-> s * [1.15] pzcmi7t}{}
\DeclareMathAlphabet{\mathpzc}{OT1}{pzc}{m}{it}

\newcommand{\be}{\begin{equation}}
\newcommand{\bea}{\begin{eqnarray}}
\newcommand{\ee}{\end{equation}}
\newcommand{\eea}{\end{eqnarray}}

\def\s#1{{\scriptscriptstyle #1}}
\def\smath#1{{\scriptscriptstyle\mathrm{#1}}}

\newcommand{\Gnp}{\Gamma}

\def\1eq#1{Eq.~(\ref{#1})}

\def\2eqs#1#2{Eqs.~(\ref{#1}) and~(\ref{#2})}
\def\3eqs#1#2#3{Eqs.~(\ref{#1}),~(\ref{#2}) and~(\ref{#3})}

\newcommand{\fatg}{{\rm{I}}\!\Gamma}

\def\ie{{\it i.e.}, }
\def\eg{{\it e.g.}, }

\def\s#1{{\scriptscriptstyle #1}}

\def\lsim{\raise0.3ex\hbox{$<$\kern-0.75em\raise-1.1ex\hbox{$\sim$}}}
\def\gsim{\raise0.3ex\hbox{$>$\kern-0.75em\raise-1.1ex\hbox{$\sim$}}}
\def\bea{\begin{eqnarray} }
\def\beq{\begin{eqnarray} }

\def\eea{\end{eqnarray}}
\def\eeq{\end{eqnarray}}

\def\assymto#1{\mbox{\raisebox{-1.2ex}[0.ex][1.6ex]{$\stackrel{\simeq}{\scriptscriptstyle #1}$}}}

\begin{document}

\title{Gluon propagator and three-gluon vertex with dynamical quarks}

\author{A.~C.~Aguilar}
\affiliation{\mbox{University of Campinas - UNICAMP, Institute of Physics ``Gleb Wataghin,''} 
13083-859 Campinas, S\~{a}o Paulo, Brazil.}
\author{F.~De Soto}
\affiliation{\mbox{Dpto. Sistemas F\'isicos, Qu\'imicos y Naturales}, \\
 Univ. Pablo de Olavide, 41013 Sevilla, Spain.}
\author{M.~N. Ferreira}
\affiliation{\mbox{University of Campinas - UNICAMP, Institute of Physics ``Gleb Wataghin,''} 
13083-859 Campinas, S\~{a}o Paulo, Brazil.}
\author{J.~Papavassiliou}
\affiliation{Department of Theoretical Physics and IFIC, University of Valencia and CSIC, E-46100, Valencia, Spain.}
\author{J.~Rodr\'{\i}guez-Quintero}
\affiliation{\mbox{Department of Integrated Sciences,  
University of Huelva, E-21071 Huelva, Spain.}}
\author{S.~Zafeiropoulos}
\affiliation{Aix Marseille Univ, Universit\'{e} de Toulon, CNRS, CPT, Marseille, France.}

\begin{abstract}

We  present  a  detailed  analysis  of  the  kinetic  and  mass  terms
associated with the  Landau gauge gluon propagator in  the presence of
dynamical  quarks,  and a  comprehensive  dynamical  study of  certain
special  kinematic limits  of  the three-gluon  vertex.  Our  approach
capitalizes on  results from  recent lattice simulations  with (2+1)
domain wall  fermions, a novel  nonlinear treatment of the  gluon mass
equation, and  the nonperturbative reconstruction of  the longitudinal
three-gluon  vertex from  its  fundamental Slavnov-Taylor  identities.
Particular emphasis  is placed  on the  persistence of  the suppression
displayed  by  certain combinations  of  the  vertex form  factors  at
intermediate  and  low  momenta,  already  known  from  numerous  pure
Yang-Mills studies.  One of our central findings is that the inclusion
of dynamical  quarks moderates the  intensity of this  phenomenon only
mildly, leaving  the asymptotic  low-momentum behavior  unaltered, but
displaces  the   characteristic  ``zero crossing''  deeper   into  the
infrared region.  In addition, the effect of the three-gluon vertex is
explored  at   the  level   of  the   renormalization-group  invariant
combination corresponding to the  effective gauge coupling, whose size
is considerably reduced with respect  to its counterpart obtained from
the ghost-gluon vertex. The main upshot of the above considerations is
the further  confirmation of  the tightly interwoven  dynamics between
the two- and three-point sectors of QCD.

\end{abstract}


\maketitle

\section{Introduction}

The three-gluon vertex of QCD~\cite{Marciano:1977su,Ball:1980ax,Davydychev:1996pb,Gracey:2011vw}, to be denoted by $\Gamma_{\alpha\mu\nu}$, has received particular attention in recent years because, in addition to its phenomenological relevance, it displays
features that are inextricably connected with subtle dynamical mechanisms operating  
in the two-point sector of the theory. In particular,
the emergence of a gluonic mass scale~\cite{Cornwall:1981zr,Bernard:1981pg,Bernard:1982my,Donoghue:1983fy,Wilson:1994fk,Philipsen:2001ip,Aguilar:2002tc}, in conjunction with the
nonperturbative masslessness of the ghost field~\cite{Alkofer:2000wg,Fischer:2006ub,Aguilar:2008xm,Boucaud:2008ky}, would appear to account   
for the ``infrared (IR) suppression'' of the basic form factors of $\Gamma_{\alpha\mu\nu}$, 
established in lattice simulations~\cite{Cucchieri:2006tf,Cucchieri:2008qm,Athenodorou:2016oyh,Duarte:2016ieu} as well as in numerous continuum approaches~\cite{
Huber:2012zj,Pelaez:2013cpa,Aguilar:2013vaa,Blum:2014gna,Blum:2015lsa,Eichmann:2014xya,Mitter:2014wpa,Williams:2015cvx,Cyrol:2016tym,Corell:2018yil,Aguilar:2019jsj}. 

The IR saturation of
the Landau gauge gluon propagator~\cite{Cucchieri:2007md,Bogolubsky:2007ud,Bogolubsky:2009dc,Oliveira:2009eh,Ayala:2012pb,Aguilar:2004sw,Aguilar:2006gr,Aguilar:2008xm,Fischer:2008uz,Boucaud:2008ky,Dudal:2008sp,RodriguezQuintero:2010wy,Tissier:2010ts,Pennington:2011xs,Cloet:2013jya,Fister:2013bh,Cyrol:2014kca,Binosi:2014aea,Aguilar:2015bud,Cyrol:2018xeq}, $\Delta(q^2)$,
has been extensively studied within  the framework developed from the
fusion of the pinch-technique
(PT)~\cite{Cornwall:1981zr,Cornwall:1989gv,Pilaftsis:1996fh,Binosi:2009qm}
with  the  background-field method
(BFM)~\cite{Abbott:1980hw},
known as the ``PT-BFM   scheme''~\cite{Aguilar:2006gr,Binosi:2007pi}.
From the dynamical point of view, the saturation is explained
by implementing the Schwinger mechanism
at the level of the SDE that controls the momentum evolution of 
$\Delta(q^2)$~\cite{Binosi:2012sj,Aguilar:2015bud}.
In this context, it is natural to regard $\Delta(q^2)$ as the sum of two distinct components,
the ``kinetic term'', $J(q^2)$, and the (momentum-dependent)
mass term, $m^2(q^2)$, as shown in \1eq{eq:gluon_m_J}.
This splitting enforces a special realization of the Slavnov-Taylor identity (STI) 
satisfied by the fully dressed $\Gamma_{\alpha\mu\nu}$~\cite{Binosi:2012sj}, 
which allows the reconstruction of its longitudinal part 
by means of a nonperturbative generalization~\cite{Aguilar:2019jsj} of the well-known Ball-Chiu (BC) construction~\cite{Ball:1980ax}. 
Specifically, the 10 longitudinal form factors of $\Gamma_{\alpha\mu\nu}$, to be denoted by $X_i$, 
are fully determined by the $J(q^2)$, the ghost dressing function, $F(q^2)$, and three of the
five form factors comprising the ghost-gluon kernel, $H_{\mu\nu}$~\cite{Ball:1980ax,Davydychev:1996pb,Aguilar:2018csq}.
However, out of all these ingredients, it is the $J(q^2)$ that is largely responsible 
for the main qualitative characteristics of the $X_i$~\cite{Aguilar:2019jsj}.

As has been explained in earlier works,
the SDE governing the $J(q^2)$ is composed by two types of
(dressed) loops, those  
containing gluons with a {\it dynamically generated mass scale},
and those with massless ghosts~\cite{Aguilar:2013vaa}. The former furnish contributions that,
due to the presence of the mass, are regulated in the IR, while the latter
give rise to {\it ``unprotected''} logarithms, of the type $\ln(q^2/\mu^2)$, which diverge as $q^2\to 0$.  
The combined effect of these terms is rather striking: 
as the (Euclidean) momentum $q^2$ decreases,  
$J(q^2)$ departs  gradually from its tree-level value (unity),   
reverses its sign (``zero crossing''), and finally diverges logarithmically at the origin~\cite{Aguilar:2013vaa}.
Quite interestingly, the same overall pattern is displayed by the special combinations
of vertex form factors studied in the (quenched) SU(2) lattice simulations of~\cite{Cucchieri:2006tf,Cucchieri:2008qm} and SU(3)~\cite{Athenodorou:2016oyh,Duarte:2016ieu,Boucaud:2018xup}, 
exposing the deep connection between the two- and three-point sectors of the theory, 
encoded in the fundamental STIs.

To date, the three-gluon vertex studies carried out 
within the PT-BFM framework  
have been limited to the pure Yang-Mills theory~\cite{Aguilar:2019jsj}. 
In the present work, 
we take a closer look at the structure of this vertex
in the presence of dynamical quarks, thus making contact with real-world QCD.

In particular, 
we present and analyze results for $\Delta(q^2)$ and $\Gamma_{\alpha\mu\nu}$ 
obtained from numerical simulations of lattice QCD, using ensembles of gauge fields with $N_f=2+1$ domain wall fermions~\cite{Blum:2014tka,Boyle:2015exm,Boyle:2017jwu}, at the physical point, \mbox{$m_\pi= 139$ MeV}.
These lattice results are complemented by a detailed analysis based on the gluon SDE and the STIs
that connect the kinetic term of $\Delta(q^2)$ with the form factors of $\Gamma_{\alpha\mu\nu}$;
for brevity, we will refer to our continuum treatment as {\it ``SDE-based''}.  
Within this latter approach, the ``unquenched'' $J(q^2)$ is determined following the procedure
first introduced in~\cite{Aguilar:2019kxz}, 
using as aid the aforementioned lattice results for $\Delta(q^2)$.
Then, the $J(q^2)$ is employed as the main ingredient of the nonperturbative
BC construction introduced in~\cite{Aguilar:2019jsj}, which provides 
definite predictions for the two special 
combinations of vertex form factors, denoted
by $\overline{\Gamma}_1^{\,\rm sym}(q^2)$ and $\overline{\Gamma}_3^{\,\rm asym}(q^2)$, 
considered in our lattice simulation. 

The main findings of this work may be summarized as follows.
{\it (i)} There is excellent agreement between the SDE-based calculation and the 
lattice data for $\overline{\Gamma}_1^{\,\rm sym}(q^2)$ and $\overline{\Gamma}_3^{\,\rm asym}(q^2)$.
{\it (ii)} Given that all quark loops are tamed in the IR by 
the constituent quark masses, the logarithmic divergence displayed by $J(q^2)$
is still controlled by the ghost-loop, which is essentially insensitive to unquenching effects~\cite{Ayala:2012pb}.
{\it (iii)} The deep IR behavior of $\overline{\Gamma}_1^{\,\rm sym}(q^2)$ and $\overline{\Gamma}_3^{\,\rm asym}(q^2)$
is determined by the corresponding asymptotic form of $J(q^2)$, multiplied by the value of the ghost dressing function
at the origin, namely $F(0)$. 
{\it (iv)} The positions of the zero crossings displayed by the unquenched $J(q^2)$,
$\overline{\Gamma}_1^{\,\rm sym}(q^2)$, and $\overline{\Gamma}_3^{\,\rm asym}(q^2)$
move deeper into the IR region with respect to the quenched cases, 
in agreement with the results reported in~\cite{Williams:2015cvx}.
{\it (v)} The suppression of $J(q^2)$, and, correspondingly, of 
$\overline{\Gamma}_1^{\,\rm sym}(q^2)$ and $\overline{\Gamma}_3^{\,\rm asym}(q^2)$,
is about 25\% milder than in the quenched case.

The article is organized as follows.
In Sec.~\ref{sec:overview} we introduce the necessary concepts and notation,
and define the quantities studied in the lattice simulation. 
In Sec.~\ref{sec:theJ} we present the salient theoretical notions associated with the gluon kinetic term,  $J(q^2)$, 
and outline the procedure that permits us its indirect determination when dynamical quarks are included. 
Next, in Sec.~\ref{sec:irsup} the SDE-based predictions for
$\overline{\Gamma}_1^{\,\rm sym}(q^2)$ and $\overline{\Gamma}_3^{\,\rm asym}(q^2)$ are derived,
and subsequently compared with the lattice results.
Moreover, the corresponding running couplings are constructed,
and directly compared with the corresponding quantity obtained from the ghost-gluon vertex.
Finally, in Sec.~\ref{sec:concl} we discuss the results and summarize our conclusions.

\section{\label{sec:overview}The three-gluon vertex: General considerations}

In this section we first present the basic definitions and conventions related with the gluon propagator and the
three-gluon vertex. Then, we review the two main quantities (vertex projections) that 
have been evaluated in the lattice simulation reported here.

\subsection{\label{sec:gen}Notation and basic properties}

Throughout this article 
we work in the {\it Landau gauge}, where the gluon propagator
is completely transverse, 
\begin{eqnarray}\label{eq:prop}
\Delta^{ab}_{\mu\nu}\left(p\right) \ = \ \langle \widetilde{A}^a_\mu(p) \widetilde{A}^b_\nu(-p) \rangle \ = \ 
\Delta(p^2) \delta^{ab}  P_{\mu \nu}(p) \ ;
\end{eqnarray}
$\widetilde{A}_{\mu}^{a}$ are the  SU(3) gauge fields in Fourier space,
the average $\langle \cdot \rangle$ indicates functional integration over the gauge space, 
and  $P_{\mu\nu}(p) = g_{\mu\nu}-p_\mu p_\nu/p^2$.

\begin{figure}[h]
\centering 
\includegraphics[scale=0.7]{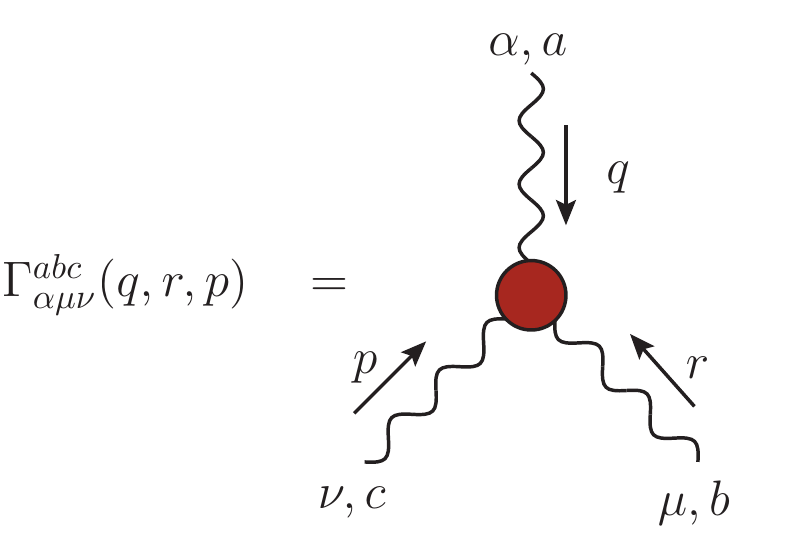}
\caption{The three-gluon vertex and the corresponding momentum/index conventions.}
\label{fig:vertex}
\end{figure}

In addition, we introduce the ghost propagator, \mbox{ $D^{ab}(q^2)= \delta^{ab}D(q^2)$}, related to its dressing function, $F(q^2)$, by 
\be
D(q^2) =\frac{iF(q^2)}{q^2}\,.
\label{eq:ghost_dressing}
\ee

Similarly, in the three-point sector of QCD, one defines the correlation function of three gauge fields,
at momenta $q$, $r$, and $p$ (with $q+r+p=0$), 
\begin{eqnarray}
{\cal G}^{abc}_{\alpha\mu\nu}\left(q,r,p\right) 
=  \ \langle \widetilde{A}_{\mu}^{a}(q) \widetilde{A}_{\nu}^{b}(r) \widetilde{A}_{\rho}^{c}(p) \rangle 
&=& f^{abc}  {\cal G}_{\alpha \mu \nu}\left(q,r,p\right)\,, 
\label{eq:3gdef}
\end{eqnarray}
where the connected three-point function ${\cal G}_{\alpha \mu \nu}(q,r,p)$ is given by 
\be
{\cal G}_{\alpha \mu \nu} (q,r,p) =  g\, \Gamma_{\alpha' \mu' \nu'} (q,r,p) P^{\alpha'}_{\alpha}(q) P^{\mu'}_{\mu}(r) P^{\nu'}_{\nu}(p)
\Delta(q^2) \Delta(r^2) \Delta(p^2)\,,
\label{eq:3gdef3}
\ee
with  $\Gnp_{\alpha\mu\nu}(q,r,p)$ denoting the conventional one-particle irreducible (1-PI) three-gluon vertex
(see Fig.~\ref{fig:vertex}).

It is customary to introduce the {\it transversally projected} vertex, $\overline{\Gamma}_{\alpha\mu\nu}(q,r,p)$, 
defined as~\cite{Cucchieri:2006tf} 
\be
\overline{\Gamma}_{\alpha\mu\nu}(q,r,p)  = \Gamma_{\alpha' \mu' \nu'} (q,r,p) P^{\alpha'}_{\alpha}(q) P^{\mu'}_{\mu}(r) P^{\nu'}_{\nu}(p)\,,
\label{transproj}
\ee
such that 
\be
{\cal G}_{\alpha \mu \nu} (q,r,p) =  g\, \overline{\Gamma}_{\alpha\mu\nu}(q,r,p)
\Delta(q^2) \Delta(r^2) \Delta(p^2)\,.
\label{eq:3gdef2}
\ee
Evidently, 
\be
q^{\alpha} \overline{\Gamma}_{\alpha\mu\nu}(q,r,p) = r^{\mu} \overline{\Gamma}_{\alpha\mu\nu}(q,r,p)
=p^{\nu}\overline{\Gamma}_{\alpha\mu\nu}(q,r,p) =0 \,.
\label{transv}
\ee

The vertex $\Gnp_{\alpha\mu\nu}(q,r,p)$ is usually decomposed into two distinct pieces, according to\mbox{~\cite{Ball:1980ax,Davydychev:1996pb,Aguilar:2019kxz}},  
\be
\Gnp^{\alpha\mu\nu}(q,r,p) = \Gamma_{\!\s L}^{\alpha\mu\nu}(q,r,p) + \Gamma_{\!\s T}^{\alpha\mu\nu}(q,r,p)\,,
\label{decomp}
\ee
where the ``longitudinal'' part, $\Gamma_{\!\s L}^{\alpha\mu\nu}(q,r,p)$, saturates the corresponding STIs [see \1eq{stig}], 
while the totally ``transverse'' part, $\Gamma_{\!\s T}^{\alpha\mu\nu}(q,r,p)$, satisfies \1eq{transv}.

The tensorial decomposition of $\Gamma_{\!\s L}^{\alpha\mu\nu}(q,r,p)$ and $\Gamma_{\!\s T}^{\alpha\mu\nu}(q,r,p)$
reads 
\be
\Gamma_{\!\s L}^{\alpha\mu\nu}(q,r,p) = \sum_{i=1}^{10} X_i(q,r,p) \ell_i^{\alpha\mu\nu} \,,
\qquad \Gamma_{\!\s T}^{\alpha\mu\nu}(q,r,p) = \sum_{i=1}^{4}Y_i(q,r,p)t_i^{\alpha\mu\nu} \,,
\label{BCbasis}
\ee
where the explicit expressions of the basis elements $\ell_i^{\alpha\mu\nu}$ and $t_i^{\alpha\mu\nu}$
are given in Eqs.~(3.4) and~(3.6) of~\cite{Aguilar:2019jsj}, respectively.

It is clear that, due to \1eq{transv}, $\overline{\Gamma}_{\alpha\mu\nu}(q,r,p)$ may be expressed entirely in terms of the 4 tensors $t_i^{\alpha\mu\nu}$, \ie 
\be\label{GammaYiXj}
\overline{\Gamma}^{\alpha\mu\nu}(q,r,p) = \sum_{i=1}^{4}\, \left[ Y_i(q,r,p) + \sum_{j=1}^{10} c_{ij} \ X_j(q,r,p) \right]
 \, t_i^{\alpha\mu\nu}\,.
\ee
The presence of the $X_j(q,r,p)$ in the final answer 
may be understood by simply noticing that, after their transverse projection, the elements   
\mbox{$\overline{\ell}_i^{\alpha\mu\nu} := \ell_i^{\alpha'\mu'\nu'}  P_{\alpha'}^{\alpha}(q) P_{\mu'}^{\mu}(r) P_{\nu'}^{\nu}(p)$},  
can be expressed as linear combinations of the $t_i^{\alpha\mu\nu}$; the exact expressions for the $c_{ij}$ may be  
straightforwardly worked out.

In addition, we define the tree-level analogue of \1eq{transproj},
\be\label{Gbartreelevel}
\overline{\Gamma}_{\!0}^{\alpha\mu\nu}(q,r,p)  =
\Gamma_{\!0}^{\alpha' \mu' \nu'} (q,r,p) P_{\alpha'}^{\alpha}(q) P_{\mu'}^{\mu}(r) P_{\nu'}^{\nu}(p)\,,
\ee
where
\be
\Gamma_{\!0}^{\alpha\mu\nu}(q,r,p) = (q-r)^{\nu}g^{\alpha\mu} + (r-p)^{\alpha}g^{\mu\nu} + (p-q)^{\mu}g^{\alpha\nu}\,.
\label{treelevel}
\ee

Note finally that, in the Euclidean space, the form factors  $X_i(q,r,p)$ and  $Y_i(q,r,p)$ are usually expressed as functions of
$q^2$,  $r^2$, and the angle $\theta$ formed between $q$ and $r$, namely $X_i(q,r,p)\to X_i(q^2,r^2,\theta)$~\cite{Aguilar:2019jsj}.

\subsection{\label{sec:lat} The lattice observables}

The lattice two- and three-point correlation functions employed in the present work 
have been obtained from $N_f$=2+1 ensembles published in~\cite{Blum:2014tka,Boyle:2015exm,Boyle:2017jwu}; 
they were generated with the Iwasaki action for the gauge sector~\cite{Iwasaki:1985we}, 
and the Domain Wall action for the fermion sector~\cite{Kaplan:1992bt,Shamir:1993zy}
(for related reviews, see, \eg~\cite{Vranas:2000tz,Kaplan:2009yg}).
In order to reach the physical point, $m_\pi = 139$ MeV, the M\"obius kernel~\cite{Brower:2004xi} has been used, resulting in a simulation of light quarks with a mass ranging from 1.3 to 1.6 MeV, while the strange quark mass is \mbox{63 MeV};
additional information on the particular setups is provided in Table~\ref{table:setup}. 
Note that the data for the gluon propagator have been recently presented in~\cite{Cui:2019dwv}, 
constituting a central ingredient in the construction of the process-independent QCD effective charge.
In addition, in an earlier work~\cite{Zafeiropoulos:2019flq}, the same data were
employed in the determination of the strong running coupling
at the $Z^0$-boson mass within the so-called Taylor scheme.
Finally, details on the Landau gauge computation of the gauge fields, and  
the correlation functions defined in \2eqs{eq:prop}{eq:3gdef}, may be found in~\cite{Ayala:2012pb,Boucaud:2017obn}. 
In addition, the treatment of the $O(4)$-breaking artifacts
has been carried out as described in~\cite{Becirevic:1999uc,Becirevic:1999hj,deSoto:2007ht,Boucaud:2018xup}.

\begin{center}
\begin{table}[h]
\begin{tabular}{|c|c|c|c|c|c|}
\hline
$\beta$ & size & $m_\pi$ [MeV] & $a^{-1}$ [GeV] & V [fm$^4$] & confs \\
\hline
2.37 & $32^3 \times 64$ & 370 & 3.148 & $2.00^3\times 4.00$ & 590 \\
\hline
2.25 & $64^3 \times 128$ & 139.15 & 2.359 & $5.35^3\times 10.70$ & 330 \\
\hline
2.13 & $48^3 \times 96$ & 139.35 & 1.730 & $5.47^3\times 10.93$ & 350 \\ 
\hline
1.63 & $48^3 \times 64$ & 137.5 & 0.997 & $9.43^3\times 12.57$ & 276 \\
\hline
\end{tabular}
\caption{Setup parameters for the four lattice ensembles used in this work.}
\label{table:setup}
\end{table}
\end{center}

Let us now consider the special quantity  
\begin{equation}
 \label{latproj}
 T(q,r,p) = \frac{W^{\alpha\mu\nu}(q,r,p){\cal G}_{\alpha\mu\nu}(q,r,p)}
{W^{\alpha\mu\nu}(q,r,p)W_{\alpha\mu\nu}(q,r,p)}\,,
\end{equation}
where the explicit form of the tensors $W_{\alpha\mu\nu}(q,r,p)$ will be judiciously chosen
in order to project out particular components of the connected three-point function, ${\cal G}_{\alpha\mu\nu}(q,r,p)$,
in certain simplified kinematic limits. Note that, in general, the quantity $T(q,r,p)$
is comprised of both longitudinal and transverse components, $X_i$ and $Y_i$.

As in~\cite{Athenodorou:2016oyh}, we focus on two special kinematic configurations: 

{\it (i)} The \emph{totally symmetric} limit, obtained when 
\be
q^2 = p^2= r^2 :=s^2\,, \quad q\cdot p = q\cdot r = p\cdot r = -\frac{s^2}{2}\,, \quad \theta=2\pi/3 \,.
\label{defsym}
\ee

{\it (ii)}
The \emph{asymmetric} limit, corresponding to the kinematic choice 
\be
p\to 0 \,,\quad r=-q\,, \quad \theta=\pi \,.
\label{defasym}
\ee

Starting with case {\it (i)}, it is relatively straightforward to establish that   
the application of the symmetric limit in \1eq{defsym} reduces the tensorial
structure of $\overline{\Gamma}^{\alpha\mu\nu}(q,r,p)$ down to~\cite{Athenodorou:2016oyh} 
\be
\overline{\Gamma}_{\rm sym}^{\alpha\mu\nu}(q,r,p) = \overline{\Gamma}^{\,\rm sym}_1 (s^2) \lambda_1^{\alpha \mu\nu}(q,r,p) +
\overline{\Gamma}_2^{\,\rm sym} (s^2)\lambda_2^{\alpha\mu\nu}(q,r,p)\,,
\label{Gsym}
\ee
with
\begin{eqnarray}
\lambda_1^{\alpha \mu\nu}(q,r,p) =  \overline{\Gamma}_{\!0}^{\alpha\mu\nu}(q,r,p)\ , \qquad
\lambda_2^{\alpha\mu\nu}(q,r,p) = \frac{(r-p)^{\alpha} (p-q)^{\mu}(q-r)^{\nu}}{s^2} \,.
\label{eq:basis}
\end{eqnarray}
The form factor  $\overline{\Gamma}_1^{\, \rm sym} (s^2)$ is particularly interesting, because it captures 
certain exceptional features linked to a vast array of underlying theoretical ideas. 
$\overline{\Gamma}_1^{\, \rm sym} (s^2)$ may be projected out by contracting \1eq{Gsym} with the tensor
\be
{\widetilde\lambda}_1^{\alpha\mu\nu} = \lambda_1^{\alpha \mu\nu}(q,r,p) + \frac{1}{2} \lambda_2^{\alpha\mu\nu}(q,r,p)\,,
\ee
which is orthogonal to $\lambda_2^{\alpha\mu\nu}(q,r,p)$.
Therefore, the substitution $W^{\alpha\mu\nu}(q,r,p) \to {\widetilde\lambda}_1^{\alpha\mu\nu}(q,r,p)$ at the level of \1eq{latproj},
and the subsequent implementation of \1eq{defsym} in the resulting expressions, leads to 

\be
T^{\rm sym}(s^2) := T(q,r,p)\,\big|_{{\rm Eq.(\ref{defsym})}}^{W \to {\widetilde\lambda}_1} = g\, \overline{\Gamma}_1^{\,\rm sym} (s^2)\Delta^3(s^2)\,.
\label{Lsym}
\ee
As has been shown in~\cite{Aguilar:2019jsj}, the use of the basis of \1eq{BCbasis} allows one to express 
$\overline{\Gamma}_1^{\,\rm sym}(s^2)$ in the form 
\be  
\label{eq:GammaSym_Xi}
\overline{\Gamma}_1^{\,\rm sym} (s^2)= X_1(s^2) - \frac{s^2}{2} X_3(s^2) + \frac{s^4}{4} Y_1(s^2) - \frac{s^2}{2} Y_4(s^2) \,.
\ee

Turning to  case {\it (ii)}, the implementation of the asymmetric limit gives rise to an expression for   
$\overline{\Gamma}^{\alpha\mu\nu}(q,r,p)$
given by a single tensor, namely~\cite{Athenodorou:2016oyh}
\be\label{Gammaasym}
\overline{\Gamma}_{\rm asym}^{\alpha\mu\nu}(q,r,p) = \overline{\Gamma}_3^{\,\rm asym} (q^2) \lambda_1^{\alpha \mu\nu}(q,-q,0)\,,
\ee
with
\be\label{lambda1asym}
\lambda_1^{\alpha\mu\nu}(q,-q,0) = 2q^\nu P^{\alpha\mu}(q) \,.
\ee

Setting $W \to \lambda_1^{\alpha\mu\nu}(q,-q,0)$ into \1eq{latproj}, one obtains
\be
T^{\rm asym}(q^2) := T(q,r,p)\,\big|_{{\rm Eq.(\ref{defasym})}}^{W \to \lambda_1} = g\overline{\Gamma}_3^{\,\rm asym}(q^2) \Delta(0)\Delta^2(q^2)\,.
\label{Lasym}
\ee
Again, using \1eq{BCbasis}, we may cast $\overline{\Gamma}_3^{\,\rm asym}(q^2)$ in the form~\cite{Aguilar:2019jsj} 
\be   
\label{eq:asyGamma}
\overline{\Gamma}_3^{\,\rm asym}(q^2) =  X_1(q^2,q^2, \pi) - q^2X_3(q^2,q^2, \pi)\, .
\ee
Interestingly, $\overline{\Gamma}_3^{\,\rm asym}(q^2)$
does not contain any reference to the transverse form factors $Y_i$, and may be therefore determined in its entirety 
by the nonperturbative BC construction of~\cite{Aguilar:2019jsj}.

\section{\label{sec:theJ}The kinetic term of the gluon propagator}

In this section we take a closer look at the kinetic term of the gluon propagator, which,
by virtue of the fundamental STIs, 
is closely connected with the longitudinal form factors $X_i$, introduced in \1eq{BCbasis}.
After reviewing certain salient theoretical concepts related to this
quantity, we outline its indirect derivation from the unquenched gluon propagator and the corresponding gluon mass equation,
and discuss some of its most outstanding properties.

\subsection{\label{subsec:concepts}Basic concepts and key relations}

A special feature of $\Delta(q^2)$, observed in the Landau gauge,
is its saturation in the deep IR~\cite{Cornwall:1981zr,Aguilar:2008xm,Aguilar:2016ock}. This property 
has been firmly established in a variety of SU(2)~\cite{Cucchieri:2007md,Cucchieri:2007rg,Cucchieri:2009zt} and SU(3)~\cite{Bogolubsky:2007ud,Boucaud:2006if,Bowman:2007du,Bogolubsky:2009dc,Oliveira:2009eh,Ayala:2012pb,Bicudo:2015rma} large-volume lattice simulations,
both quenched and unquenched. Due to its far reaching theoretical implication, this property   
has been scrutinized in the continuum using a multitude of distinct approaches~\mbox{\cite{Aguilar:2004sw,Aguilar:2006gr,Braun:2007bx,Epple:2007ut,Aguilar:2008xm,Fischer:2008uz,Boucaud:2008ky,Dudal:2008sp,RodriguezQuintero:2010wy,Tissier:2010ts,Pennington:2011xs,Cloet:2013jya,Serreau:2012cg,Aguilar:2015bud,Fister:2013bh,Binosi:2014aea,Cyrol:2014kca,Cyrol:2018xeq}.}

This characteristic behavior of $\Delta(q^2)$ is considered to be intimately connected
with the emergence of a gluon mass scale~\cite{Cornwall:1981zr,Philipsen:2001ip,Aguilar:2002tc}, and has been studied in detail 
within  the framework of the ``PT-BFM''~\cite{Aguilar:2006gr,Binosi:2007pi}.
For the purposes of the present work, we will briefly comment on a limited number of 
concepts and ingredients related with this particular approach; for further details, the reader is referred to
the extended literature cited above.

{\it (a)} The IR finiteness of $\Delta(q^2)$ motivates the splitting of its inverse 
into two separate components, according to (Euclidean space)~\cite{Binosi:2012sj} 
\be
\label{eq:gluon_m_J}
\Delta^{-1}(q^2) = q^2J(q^2) + m^2(q^2)\,,
\ee
where $J(q^2)$ corresponds to the so-called ``kinetic term''
[at tree-level, $J(q^2) =1$], 
while $m^2(q^2)$ to a momentum-dependent gluon mass scale,
with the property $m^2(0)=\Delta^{-1}(0)$.
Note that we have suppressed the dependence of all quantities appearing in \1eq{eq:gluon_m_J}
on the renormalization point $\mu$. 
For large values of $q^2$, the component $J(q^2)$ captures the 
standard perturbative corrections to the gluon propagator, while in the IR it 
exhibits exceptional nonperturbative features~\cite{Aguilar:2013vaa,Aguilar:2019jsj}.

{\it (b)} The emergence of the component $m^2(q^2)$ is triggered by the
non-Abelian realization of the well-known Schwinger mechanism~\cite{Schwinger:1962tn,Schwinger:1962tp} for gauge boson mass generation.
This latter mechanism is activated through the inclusion of 
{\it longitudinally coupled massless poles} 
into the three-gluon vertex that enters in the SDE governing the evolution of $\Delta^{-1}(q^2)$~\cite{Aguilar:2011xe,Binosi:2012sj,Ibanez:2012zk,Binosi:2017rwj,Aguilar:2017dco}.
In particular, one implements the replacement
\be
\Gnp_{\alpha\mu\nu} \to \fatg_{\alpha\mu\nu}= \Gnp_{\alpha\mu\nu} + V_{\alpha\mu\nu}\,,
\label{twovert}
\ee
where $V_{\alpha\mu\nu}$ contains the aforementioned poles, arranged in the special tensorial structure~\cite{Ibanez:2012zk} 
\be
V_{\alpha\mu\nu}(q,r,p) = \left(\frac{q_{\alpha}}{q^2}\right)A_{\mu\nu}(q,r,p) +
\left(\frac{r_{\mu}}{r^2}\right)B_{\alpha\nu}(q,r,p) + \left(\frac{p_{\nu}}{p^2}\right)C_{\alpha\mu}(q,r,p)\,.
\label{longcoupl}
\ee
Consequently, by virtue of the relation $V_{\alpha' \mu' \nu'} (q,r,p) P^{\alpha'}_{\alpha}(q) P^{\mu'}_{\mu}(r) P^{\nu'}_{\nu}(p)=0$,
the component $V_{\alpha\mu\nu}(q,r,p)$ drops out from the quantity $T(q,r,p)$ defined in \1eq{latproj},
and only the ``no-pole'' part of the vertex, $\Gnp_{\alpha\mu\nu}$, contributes to it.

{\it (c)}
It turns out that 
the two functions composing $\Delta^{-1}(q^2)$ in \1eq{eq:gluon_m_J} 
and the two vertices comprising $\fatg_{\alpha\mu\nu}$ in \1eq{twovert} are firmly linked. 
Specifically, the STI satisfied by $\fatg_{\alpha\mu\nu}(q,r,p)$,
\be
q^\alpha\fatg_{\alpha\mu\nu}(q,r,p) = F(q^2)[\Delta^{-1}(p^2) P^{\alpha}_\nu(p)H_{\alpha\mu}(p,q,r) - \Delta^{-1}(r^2)P^{\alpha}_\mu(r)H_{\alpha\nu}(r,q,p)]\,,
\ee
is naturally separated into two ``partial'' ones, 
relating the divergences of $\Gamma_{\alpha\mu\nu}$ and $V_{\alpha\mu\nu}$ with $J(q^2)$ and $m^2(q^2)$, respectively, namely
\footnote{Exactly analogous relations hold for the STIs with respect to the other two legs.}
\bea
q^\alpha\Gnp_{\alpha\mu\nu}(q,r,p) &=& F(q^2)[p^2J(p^2) P^{\alpha}_\nu(p)H_{\alpha\mu}(p,q,r) - r^2J(r^2)P^{\alpha}_\mu(r)H_{\alpha\nu}(r,q,p)] \,,
\label{stig}\\
q^\alpha V_{\alpha\mu\nu}(q,r,p) &=& F(q^2)[m^2(r^2)P^{\alpha}_\mu(r)H_{\alpha\nu}(r,q,p) - m^2(p^2) P^{\alpha}_\nu(p)H_{\alpha\mu}(p,q,r)] \,.
\label{stiv}
\eea
The practical implication of this separation is that the form factors $X_i$ of  
$\Gamma_{\!\s L}^{\alpha\mu\nu}(q,r,p)$ may be reconstructed by means of a nonperturbative
generalization~\cite{Aguilar:2019jsj} of the well-known BC procedure~\cite{Ball:1980ax}.
In particular, the $X_i$ are expressed as combinations of the $J(q^2)$, the
ghost dressing function, $F(q^2)$, and three of the five components appearing in the tensorial decomposition
of $H_{\mu\nu}$, whose one-loop dressed approximation has been computed in~\cite{Aguilar:2018csq}. These results are especially relevant for the study in hand, because
they provide a theoretical (albeit approximate) handle
on the form of the $X_i$ appearing in \2eqs{eq:GammaSym_Xi}{eq:asyGamma}; note, however, that the $Y_i$ remain
undetermined by this procedure.

{\it (d)}  The special realization of the STIs explained in point {\it (c)} 
leads to the separation of the original SDE governing 
$\Delta(q^2)$ into a system of two coupled integral equations, one determining $J(q^2)$ and the other $m^2(q^2)$~\cite{Binosi:2012sj}.  
As has been demonstrated recently in~\cite{Aguilar:2019kxz}, the self-consistent treatment
of the equation controlling $m^2(q^2)$, 
in conjunction with the (quenched) lattice data for $\Delta(q^2)$, permits one to pin down the 
form of $J(q^2)$ quite accurately, without actually invoking its own (considerably more complicated) integral equation. 
The subsequent use of this $J(q^2)$ as ingredient in the 
BC construction of the $X_i$ described above, allows one to obtain, through \2eqs{eq:GammaSym_Xi}{eq:asyGamma}, SDE-derived predictions for 
$T^{\rm sym}(s^2)$ and $T^{\rm asym}(q^2)$, which are in excellent agreement
with the lattice data of~\cite{Athenodorou:2016oyh}. 

\subsection{\label{subsec:const}The ``unquenched'' $J(q^2)$: general construction and main results}

The above considerations, and in particular the procedure summarized in point {\it (d)}, 
will be applied in the present work in order to obtain SDE-derived predictions
for the {\it unquenched} $\overline{\Gamma}^{\,\rm sym}_1 (s^2)$ and $\overline{\Gamma}_3^{\,\rm asym}(q^2)$,
which will be subsequently compared with the corresponding sets of lattice results. 
In what follows we outline the main points of this construction, postponing the 
multitude of technical details for a future communication.

($\rm P_1$): The starting point is the gluon mass equation considered in~\cite{Aguilar:2019kxz}, whose general form is given by (\mbox{$\alpha_s:= g^2/4\pi$})
\be
m^{2}(q^2) = \int \frac{\mathrm{d}^4 k}{(2\pi)^4}\, m^2(k) \Delta(k)\Delta(k+q){\cal K} (k,q,\alpha_s),
\label{meqgen}
\ee
where the kernel ${\cal K}$ receives one-loop and two-loop dressed contributions. 

($\rm P_2$):
The effective treatment of multiplicative renormalization amounts to the substitution of the 
vertex renormalization constants, multiplying the  one- and two-loop components of  ${\cal K}$, by 
kinematically simplified form factors of the three- and four-gluon vertices,
denoted by ${\cal C}_3(k^2)$ and ${\cal C}_4(k^2)$, respectively.

($\rm P_3$):
The kinetic term $J(q^2)$ enters into the gluon mass equation when the substitution given in \1eq{eq:gluon_m_J}
is implemented at the level of the term $\Delta(k)\Delta(k+q)$. In addition, the function ${\cal C}_3(k^2)$ 
depends on $J(k^2)$; specifically, for its derivation we adopt the Abelian version 
of the BC construction~\cite{Ball:1980ax}, setting the ghost dressing function and the ghost-gluon kernel at their tree-level values,
which yields simply ${\cal C}_3(k^2) = J(k^2)$.  

($\rm P_4$):
The term ${\cal C}_4(k^2)$ is approximated by the same functional form given in Eq.~(4.8) of~\cite{Aguilar:2019kxz}.
As explained there, the main feature of 
${\cal C}_4(k^2)$, which is instrumental for the stability of the gluon mass equation, is its mild enhancement
with respect to its tree-level value in the critical region of a few hundred  MeV.

\begin{figure}[t]
\begin{minipage}[b]{0.45\linewidth}
\hspace{-1.5cm}
\centering
\includegraphics[scale=0.27]{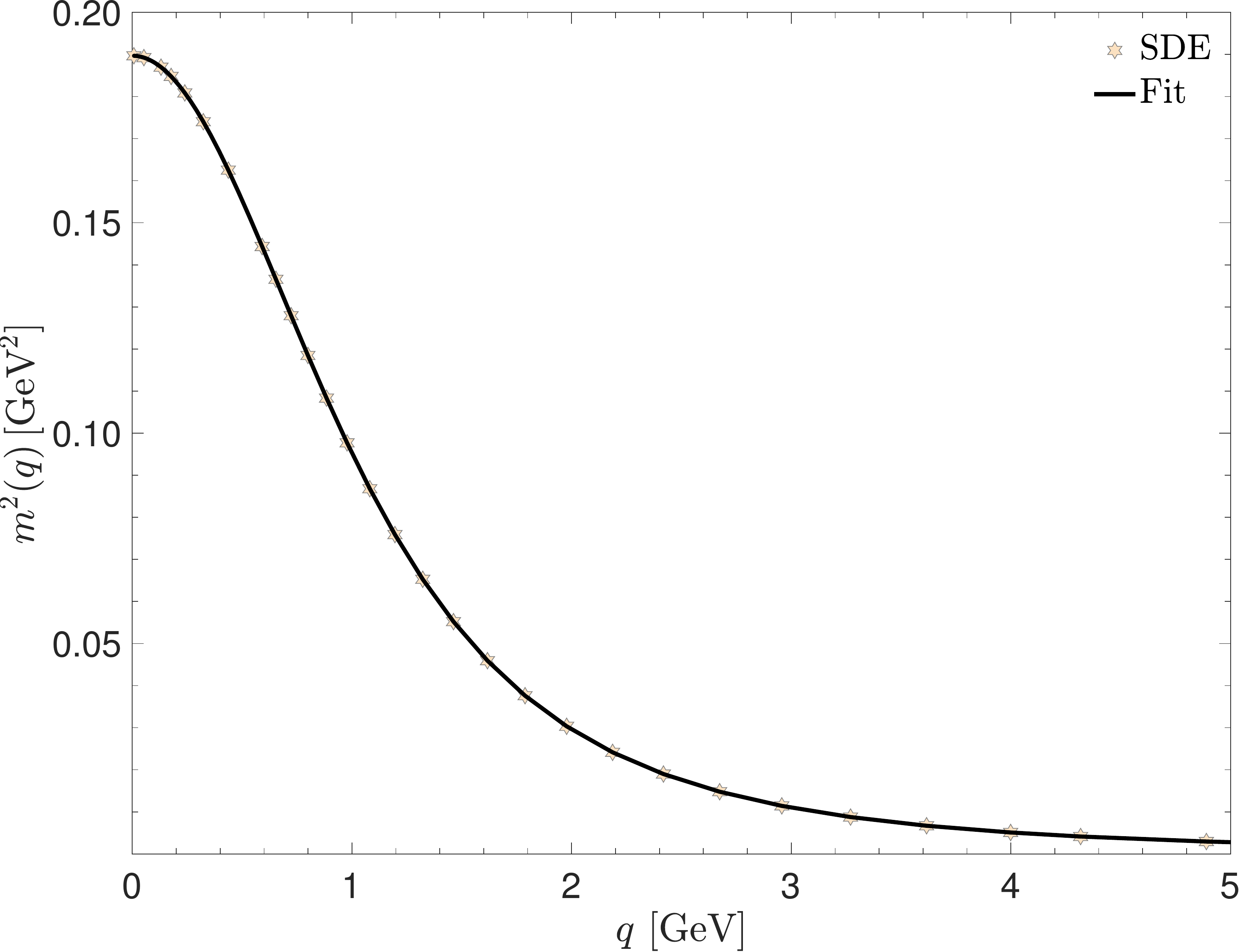}
\end{minipage}
\hspace{0.15cm}
\begin{minipage}[b]{0.45\linewidth}
\includegraphics[scale=0.27]{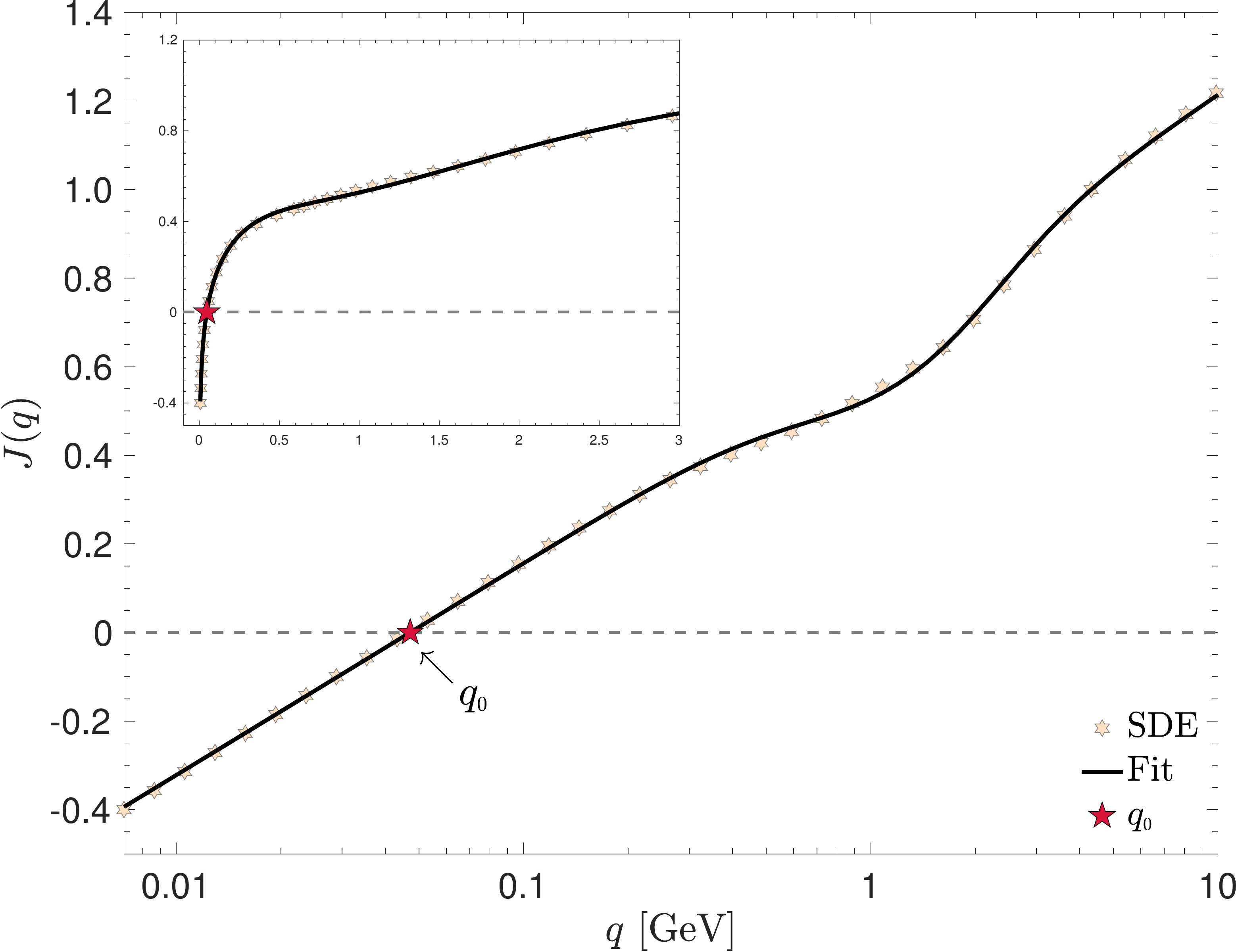}
\end{minipage}
\begin{minipage}[b]{0.45\linewidth}
\hspace{-1.5cm}
\includegraphics[scale=0.27]{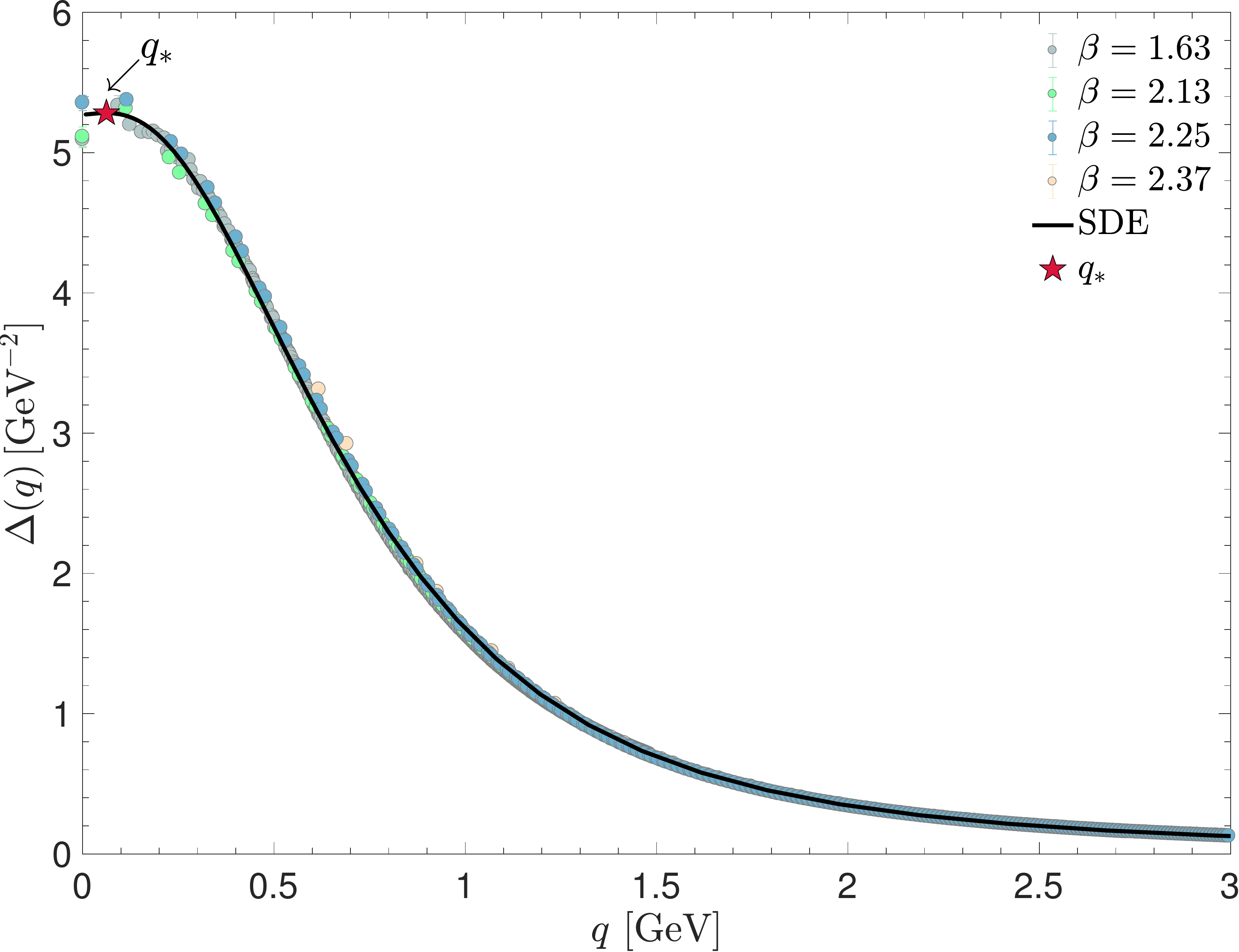}
\end{minipage}
\begin{minipage}[b]{0.45\linewidth}
\includegraphics[scale=0.27]{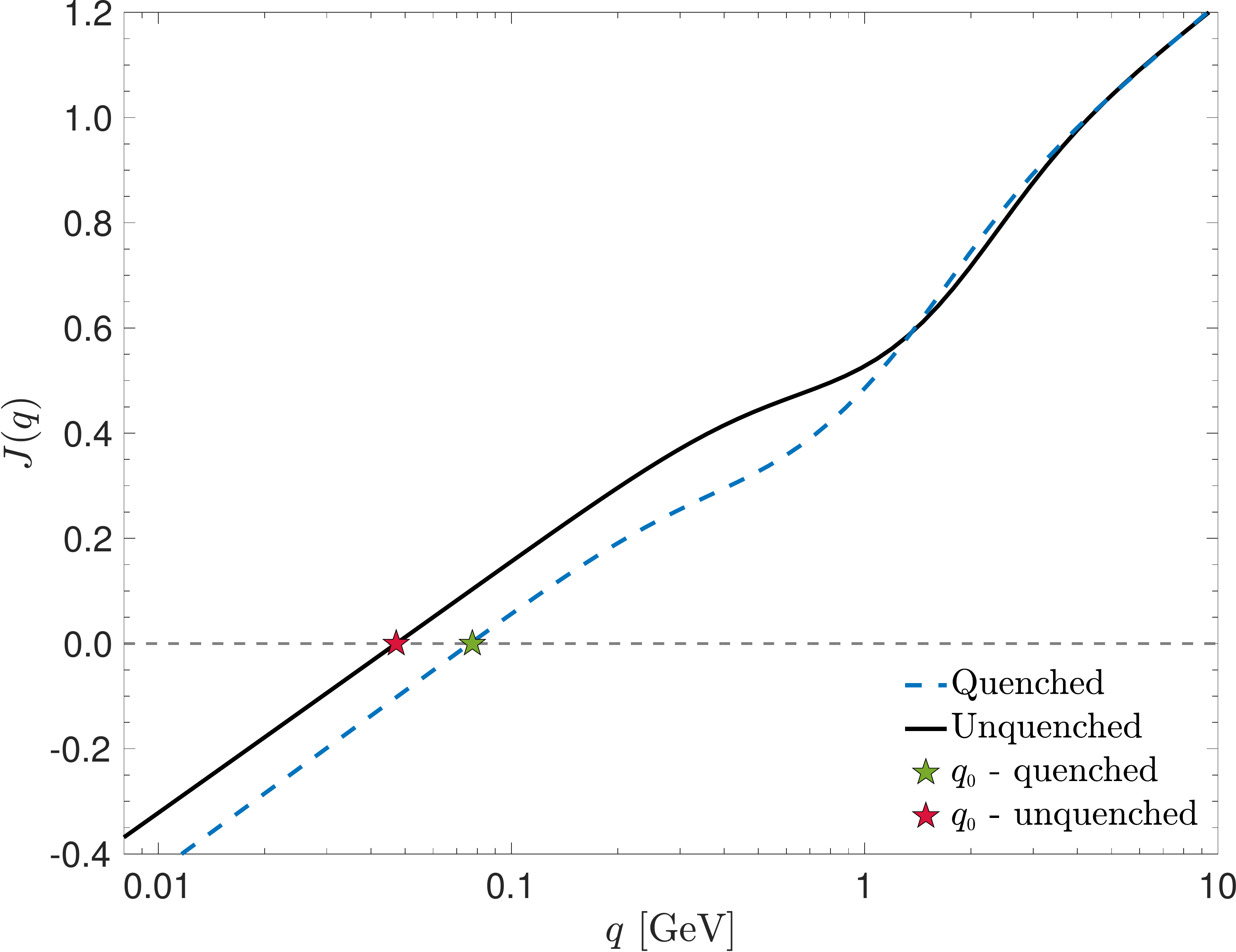}
\end{minipage}
\caption{Top left panel: The dynamical gluon mass, $m^2(q)$, obtained from Eq.~\eqref{meqgen} and fitted by Eq.~\eqref{munq}. Top right panel: The $J(q)$ obtained through the procedure described in points {\rm ($P_1$)}-{\rm ($P_5$)}, and the corresponding fit, given in~\1eq{Junq}. Bottom left panel: Comparison of the unquenched gluon propagator, $\Delta(q)$, obtained from Eq.~\eqref{eq:gluon_m_J} (black continuous),  with the lattice data (solid circles in different colors for each $\beta$). Bottom right panel: The quenched (blue dashed) and unquenched (black continuous) $J(q)$; the stars indicate the momentum 
\mbox{$q_{\scriptscriptstyle\mathrm{0}}$} where $J(q_{\scriptscriptstyle\mathrm{0}})=0$.}
\label{fig:JMDelta}
\end{figure}

($\rm P_5$):
An initial {\it Ansatz} for $J(q^2)$ is introduced as a ``seed'', and is subsequently improved by means of a well-defined
iterative procedure, described in detail in Sec.~VB of~\cite{Aguilar:2019kxz}.
In particular, both the  form of $J(q^2)$ and the value of $\alpha_s$ are
gradually modified, and each time
the corresponding solution, $m^2(q^2)$, obtained from the gluon mass equations, is recorded. 
The procedure terminates when the pair $\{m^2(q^2),J(q^2)\}$ has been identified which, 
when combined according to \1eq{eq:gluon_m_J}, provides the best possible coincidence with the lattice data for $\Delta(q^2)$ with $N_f=2+1$ [see the left bottom panel of Fig.~\ref{fig:JMDelta}].
The final value of the gauge charge is $\alpha_s=0.27$.

An excellent fit for $m^2(q^2)$, shown in the top left panel of Fig.~\ref{fig:JMDelta}, is given by 
\be 
m^2(q^2) = \frac{m_\smath{0}^4}{\kappa_\smath{1}^2 + q^2 \ln[ ( q^2 + \kappa_\smath{2}^2 )/\sigma^2 ] }  \,,
\label{munq}
\ee
where the parameters are given by \mbox{$m_\smath{0}^4 = 0.134$ GeV$^4$}, 
\mbox{$\kappa_\smath{1}^2 = 0.705$ GeV$^2$}, \mbox{$\kappa_\smath{2}^2 = 9.31$ GeV$^2$}, and
\mbox{$\sigma^2 = 5.13$ GeV$^2$}.

Similarly, the solution for  $J(q^2)$, shown in the top right panel of Fig.~\ref{fig:JMDelta}, is accurately fitted by  
\be 
J(q^2) = 1 + \frac{3\lambda_s}{4\pi}\left( 1 + \frac{\tau_\smath{1}}{q^2 + \tau_\smath{2}} \right) \left[ 2 \ln\left( \frac{q^2 + \eta^2(q^2) }{\mu^2 + \eta^2(\mu^2) } \right) + \frac{1}{6}\ln\left(\frac{q^2}{\mu^2}\right)\right] \,,
\label{Junq}
\ee
with 
\be 
\eta^2(q^2) = \frac{\eta_\smath{1}}{q^2 + \eta_\smath{2}} \,,
\label{massJ}
\ee
where $\lambda_s = 0.237$, \mbox{$\tau_\smath{1} = 7.06$ 
GeV$^2$}, \mbox{$\tau_\smath{2} = 0.709$ GeV$^2$}, \mbox{$\eta_\smath{1} = 22.35$ GeV$^4$}, 
\mbox{$\eta_\smath{2} = 1.19$ GeV$^2$}, and \mbox{$\mu^2 = 18.64$ GeV$^2$}. Notice that $J(\mu^2) = 1$, as required by the momentum subtraction (MOM) renormalization
prescription.

We emphasize that, even though several aspects of the unquenched gluon propagator have been previously addressed
within the PT-BFM formalism\footnote{For related works, see also, \eg \cite{Braun:2014ata,Fischer:2005en,Williams:2015cvx,Cyrol:2017ewj}.}~\cite{Aguilar:2012rz,Aguilar:2013hoa}, the results presented in \2eqs{munq}{Junq} are completely new. 

\subsection{\label{subsec:asymp}Asymptotic analysis for the deep IR}

By expanding the above fits for $J(q^2)$ and $m^2(q^2)$ around $q^2 \to 0$, we obtain
\be
J(q^2) = a\ln\left( \frac{q^2}{\mu^2} \right) + b \,, \qquad 
m^2(q^2) = d + c \, q^2 \,,
\label{Jmas}
\ee
and therefore
\be 
\Delta^{-1}(q^2) = d + q^2\left[ a \ln\left( \frac{q^2}{\mu^2} \right) + b + c \right]\,,
\label{Delta_IR}
\ee
with 

\bea
\displaystyle a =& \displaystyle \frac{\lambda_s}{8\pi}\left( 1 + \frac{\tau_\smath{1}}{\tau_\smath{2}} \right)\,,
\qquad b =& \displaystyle 1 + \frac{3\lambda_s}{2\pi}\left( 1 + \frac{\tau_\smath{1}}{\tau_\smath{2}} \right) \ln \left[\frac{\eta_\smath{1}}{\eta_\smath{2} \left[ \mu^2 + \eta^2(\mu^2) \right] } \right] \,,
\nonumber\\ 
& & \nonumber\\  
\displaystyle c =& \displaystyle - \frac{m_\smath{0}^4}{\kappa_\smath{1}^4} \ln\left( \frac{\kappa_\smath{2}^2}{\sigma^2}\right) \,,   
\qquad d =& \displaystyle \frac{m_\smath{0}^4}{\kappa_\smath{1}^2} \,.
\label{consabcd}
\eea

Employing the numerical values of the parameters in Eqs.\eqref{Junq} and~\eqref{massJ}, one obtains 
\mbox{$a = 0.104$}, \mbox{$b = 0.934$}, \mbox{$c = -0.160$}, \mbox{$d = 0.190$ GeV$^2$}. 

With the above asymptotic expressions at our disposal, we proceed
to elaborate on the following important points. 

{\it (i)}
As can be seen in the bottom right panel of Fig.~\ref{fig:JMDelta}, for momenta lower than about $500$ MeV,  
the quenched and unquenched $J(q^2)$ run nearly parallel to each other. 
In view of \1eq{Jmas}, this indicates that the coefficient of the logarithm remains practically
unchanged in the presence of quark loops, whose net effect in the deep IR is to simply
modify (increase) the numerical value of the constant $b$, thus shifting the position of the
zero crossing towards lower momenta. A qualitative explanation of these
observations may be given by noting
that {\it (a)} the ghost dressing function is rather insensitive to 
unquenching effects~\cite{Ayala:2012pb},
and hence, the contribution of the ghost loops is essentially the same, and
{\it (b)} the quark loops provide IR finite contributions, since the corresponding logarithms are
protected by the quark masses; their size and sign is consistent with the
analysis presented in~\cite{Aguilar:2012rz}.
It is important to emphasize, however, that  
throughout our present derivation, no quark loops have been actually evaluated; 
instead, by means of the optimization procedure described in ($\rm P_5$),
the effects of the dynamical quarks, implicit in the lattice data for $\Delta(q^2)$, 
have been {\it indirectly transmitted} to the individual components $J(q^2)$ and $m^2(q^2)$.

{\it (ii)} From \1eq{Jmas} we can obtain a particularly accurate estimate of the position of the ``zero crossing'',
\ie the momentum $q_{\scriptscriptstyle\mathrm{0}}$ for which 
\mbox{$J(q_{\scriptscriptstyle\mathrm{0}}) = 0$}; it is given by 
\be 
q_{\scriptscriptstyle\mathrm{0}} = \mu \, e^{-\frac{b}{2a} } \,.
\ee
With the values of the coefficients found before, this leads to \mbox{$q_{\scriptscriptstyle\mathrm{0}} = 48.19$ MeV}. On the other hand, computing the crossing of the full fit of \1eq{Junq} numerically yields \mbox{$q_{\scriptscriptstyle\mathrm{0}} = 47.18$ MeV} [see 
the red star in the bottom right panel of Fig.~\ref{fig:JMDelta}]. Thus, the asymptotic form is accurate to within $0.3\%$ for the position of the crossing of $J(q^2)$.

{\it (iii)} Let us next consider the maximum of $\Delta(q^2)$, and denote by $q_{*}$ the momentum where it occurs,
namely the solution of the condition $\Delta^{\prime}(q^2)=0$,
where the ``prime'' denotes differentiation with respect to $q^2$.
The appearance of this maximum is inextricably connected
with the presence of the unprotected logarithm originating from the ghost loop.
In addition to confirming the known nonperturbative behavior  
of the ghost propagator in Euclidean space (\ie absence of a ``ghost mass''),
it has a  
direct implication on the general analytic structure of the gluon propagator~\cite{Cyrol:2018xeq,Kern:2019nzx}. 
In particular, from the standard K\"all\'en-Lehmann representation~\cite{Kallen:1952zz,Lehmann:1954xi}  
\be
\Delta (q^2) = \int_0^{\infty} \!\! d t \, \frac{\rho (t)}{q^2 + t}\,,
\label{spec}
\ee
where  $\rho\, (t)$ is the gluon spectral function, we have that  
\be
\Delta^{\prime}(q^2) =
- \int_0^{\infty} \!\! d t \, \frac{\rho (t)}{(q^2+ t)^2}\,.
\label{derspec}
\ee
Then, the maximum for $\Delta (q^2)$ at $q^2=q^2_{*}$
leads necessarily to {\it positivity violation}~\cite{Osterwalder:1973dx,Osterwalder:1974tc,Alkofer:2000wg,Cornwall:2013zra}, because the condition    
\be
\int_0^{\infty} \!\! d t \, \frac{\rho (t)}{(q^2_{*} + t)^2} = 0\,,  
\label{specmax}
\ee
may be fulfilled only if $\rho(t)$ is not positive-definite.

A reasonable estimate for the value of $q_{*}$ may be derived from \1eq{Delta_IR}; specifically  
one obtains the equation  
\be 
[\Delta^{-1}(q^2)]^{\prime} = a \ln\left( \frac{q^2}{\mu^2} \right) + {\tilde c}  = 0 \,,
\ee
where ${\tilde c} := a+b+c$,  whose solution is given by 
\be 
q_{*} = \mu \exp\left( - \frac{{\tilde c}}{ 2 a }\right) \,, 
\label{Max_pos}
\ee
yielding the numerical value \mbox{$q_{*} = 63$ MeV}.

The expression for the gluon propagator at the maximum is given by
\be 
\Delta_{*} := \Delta(q^2_{*}) =
\left[d - a \mu^2 \exp\left( - \frac{{\tilde c}}{ a }\right) \right]^{-1} \,,
\label{Max_val}
\ee
its numerical value is given by $\Delta_{*} = 5.28$ GeV$^{-2}$.

{\it (iv)} Finally, we turn to another characteristic feature associated with the presence of the unprotected logarithm,
namely the logarithmic divergence of $\Delta^{\prime}(q^2)$ at the origin.
In particular, using \1eq{Delta_IR}, it is straightforward to establish that  
\be
\Delta^{\prime}(q^2) \assymto{q^2 \to 0} -\frac{a}{d^2}\ln\left( \frac{q^2}{\mu^2} \right) \to + \infty \,,
\qquad [\Delta^{-1}(q^2)]^{\prime} \assymto{q^2 \to 0} a \ln\left( \frac{q^2}{\mu^2} \right) \to - \infty \,.
\ee

\begin{figure}[t]
\centering 
\includegraphics[scale=0.27]{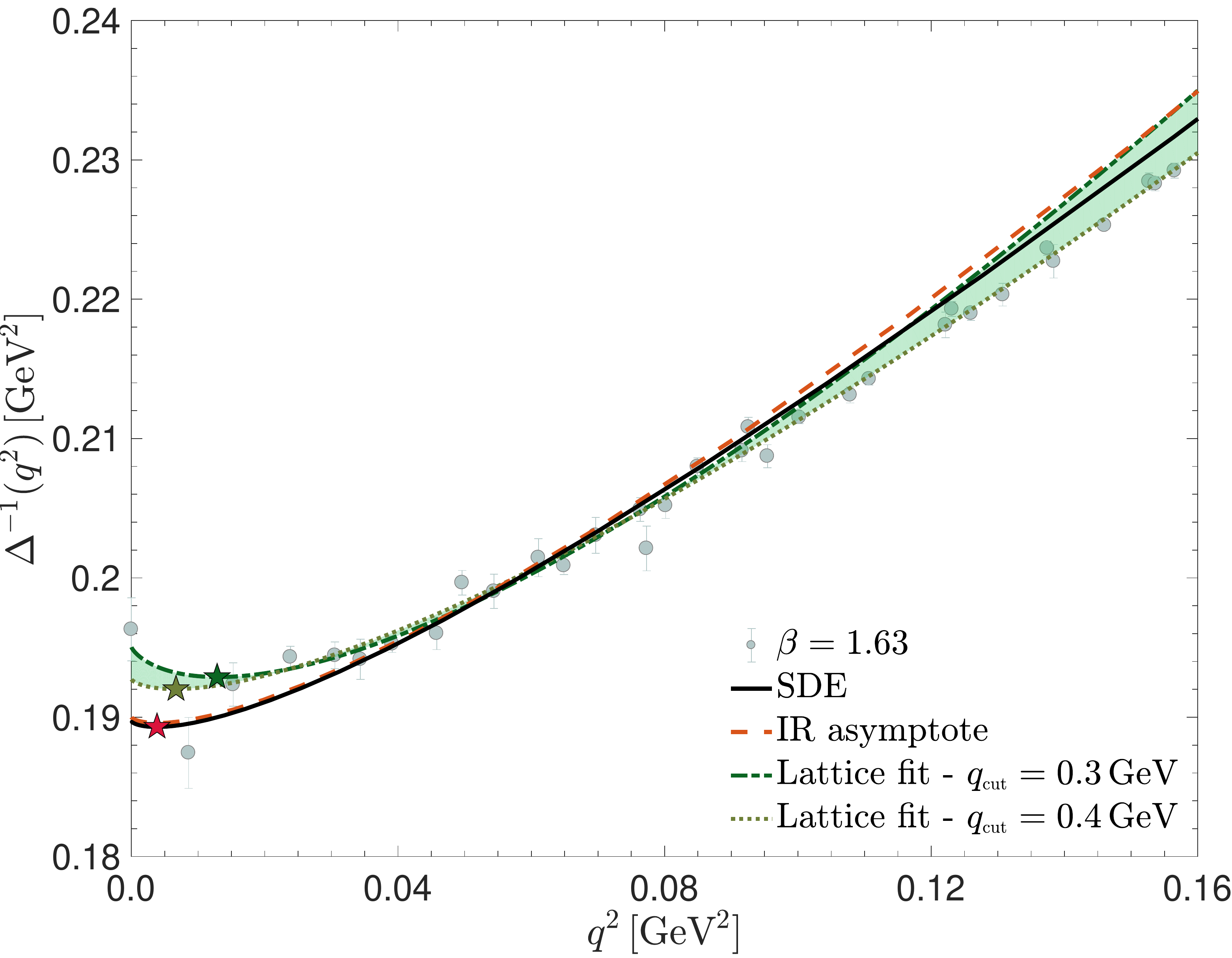}
\caption{The green band represents the asymptotic fits for the lattice data (gray solid  circles) with \mbox{$q_{{\scriptscriptstyle\mathrm{cut}}} =0.3$ GeV} (green dash-dotted line) and \mbox{$q_{\scriptscriptstyle\mathrm{cut}} =0.4$ GeV} (green dashed curve), given by 
Eq.~\eqref{Delta_IR}. The black continuous line corresponds to the SDE-based result, while the red dashed curve is its asymptotic limit.  All asymptotic curves are given by Eq.~\eqref{Delta_IR}, with the corresponding fitting parameters listed in the Table~\ref{tab:fits}.}
\label{fig:fits}
\end{figure}

{\it (v)} While the functional form of $\Delta^{-1}(q^2)$ is motivated by sound theoretical considerations, 
the numerical values for the parameters $a$, $b$, $c$, and $d$, quoted below \1eq{consabcd},
have been obtained by fitting the {\it entire range} of the SDE solution.
It would be therefore interesting to probe the stability of our asymptotic results by
contrasting them directly with the 
low-momentum domain of the lattice data, and subsequently refitting the aforementioned parameters. 
To that end, we consider only the lattice ensemble with $\beta$=1.63, because it 
contains the largest number of points in the desired region. 
Our fitting procedure is limited 
to the data below a given momentum cutoff, 
$q_{\scriptscriptstyle\mathrm{cut}}$;  we have chosen two 
values for it, namely \mbox{$q_{\scriptscriptstyle \mathrm{cut}}$=0.3 GeV} and \mbox{$q_{\scriptscriptstyle \mathrm{cut}}$=0.4 GeV}. 
The result of this analysis can be found in Fig.~\ref{fig:fits} and in Table~\ref{tab:fits}.

The  black continuous line corresponds to the SDE-based result, while the red dashed curve is its asymptotic limit.  All asymptotic curves are obtained with Eq.~\eqref{Delta_IR} using the fitting parameters  listed in the Table~\ref{tab:fits}.

\begin{center}
\begin{table}
\begin{tabular}{|c|c|c|c|c|}
\hline
IR asymptotic fits &  $a$  & $b+c$ & \,$d$ [GeV$^2$] & \, $q_*$ [MeV] \\
\hline
Lattice - ${q_{\scriptscriptstyle \mathrm{cut}}=0.3}$ {[GeV]} & \quad 0.165 \quad  &\quad 1.036 \quad  &  0.195 & 113\\
\hline
Lattice - ${q_{ \scriptscriptstyle\mathrm{cut}}=0.4}$ {[GeV]} & \quad  0.107 \quad   & \quad  0.746 \quad   & 0.193 &  82\\
\hline
\mbox{SDE expansion }& \quad  0.104 \quad  &\quad  0.774 \quad & 0.190 & 63 \\
\hline
\end{tabular}
\caption{ Fitting parameters of \1eq{Delta_IR} for different values of the cutoff $q_{ \scriptscriptstyle \mathrm{cut} }$ considered in the fitting process of the lattice data with $\beta=1.63$. In the last line, we show the fitting parameters of the SDE asymptotic limit, given by \1eq{Delta_IR}, which was obtained by expanding the result given in Eqs.~\eqref{munq} and~\eqref{Junq}. In the last column, we quote the momentum $q_*$, where the minimum of $\Delta^{-1}(q^2)$ [or maximum of $\Delta(q^2)$] occurs for each case.}
\label{tab:fits}
\end{table}
\end{center}

As we can see in Fig.~\ref{fig:fits}, the asymptotic expression  
of \1eq{Delta_IR} describes the lattice data particularly well.
Essentially, the difference between the asymptotic limit of the SDE result (red dashed line) and the best fits for the IR lattice points 
(green band) appears for very low momenta, and is of the order of $3\%$. 
The lattice data for $\Delta^{-1}(q^2)$ show a linear behavior, consistent with a $q^2$-increase, 
except for momenta below 180  MeV, where  the effect of  the logarithm  in \1eq{Delta_IR} becomes  apparent.
Note also the onset of a steep derivative close to the origin, in qualitative agreement with point {\it (iv)}.  
In addition, the refitted values of $a$, $b+c$, and  $d$
are completely consistent with those obtained from the full-range fit of the SDE result.

\section{\label{sec:irsup} IR suppression of the three-gluon vertex}

In this section we present the SDE-based computation of 
$\overline{\Gamma}^{\,\rm sym}_1 (s^2)$ and $\overline{\Gamma}_3^{\,\rm asym}(q^2)$.  
After an instructive study of the low-momentum limit,
our results for the entire range of momenta are presented and compared 
with the new lattice data. In addition, the two effective couplings obtained from 
$\overline{\Gamma}^{\,\rm sym}_1 (s^2)$ and $\overline{\Gamma}_3^{\,\rm asym}(q^2)$ are constructed,
and the former is compared with the corresponding quantity obtained from the ghost-gluon vertex.

\subsection{\label{sec:gensde} The SDE-based derivation}

The detailed form of the function $J(q^2)$ captured by \1eq{Junq} constitutes a key ingredient for the approximate evaluation of the vertex form factors
$\overline{\Gamma}^{\,\rm sym}_1 (s^2)$ and $\overline{\Gamma}_3^{\,\rm asym}(q^2)$
by means of the main equations \1eq{eq:GammaSym_Xi} and \1eq{eq:asyGamma}, respectively. 
This becomes possible because the nonperturbative BC construction
of~\cite{Aguilar:2019jsj} allows one to express the $X_i$ in terms of the kinetic term of the gluon propagator,
the ghost dressing function, and the ghost-gluon form factors. 
Even though this procedure does not determine the terms 
\mbox{$(s^4/{4}) Y_1(s^2) - ({s^2}/{2}) Y_4(s^2)$} contributing to $\overline{\Gamma}^{\,\rm sym}_1 (s^2)$, 
the overall agreement with the (admittedly error-burdened) lattice results suggests that their omission does not alter drastically
the {\it qualitative} features of the BC solution; see also the related discussions in Sec.~\ref{sec:comp}. 

Focusing precisely on  $\overline{\Gamma}_1^{\,\rm sym} (s^2)$,  
the part that depends on the two longitudinal components is given by~\cite{Aguilar:2019jsj} 
\be 
X_1(s^2) - \frac{s^2}{2} X_3(s^2)= F(s^2)\left[ J(s^2) \left( H_1(s^2) + \frac{s^2}{2} H_3(s^2) \right) +
\frac{s^2}{2} \,\frac{d J(s^2)}{ds^2} \, H_2(s^2) \right]
\,,
\label{BCstuff}
\ee
where  
\bea
H_1(s^2) &=& A_1(s^2) - \frac{s^2}{2} A_3(s^2)  \,, \nonumber\\
H_2(s^2) &=& A_1(s^2) + \frac{s^2}{2}\left[A_3(s^2) - A_4(s^2) \right] \,, \nonumber\\
H_3(s^2) &=& A_1^{(1,0,0)}(s^2) + \frac{ \sqrt{3} }{2s^2}A_1^{(0,0,1)}(s^2) + \frac{s^2}{2}\left[A_3^{(1,0,0)}(s^2) - A_4^{(1,0,0)}(s^2)\right]
\label{theH}
\nonumber\\
&+& \frac{ \sqrt{3} }{4}\left[ A_3^{(0,0,1)}(s^2) - A_4^{(0,0,1)}(s^2)\right] \,,
\eea
with the partial derivatives defined as
\begin{align}
A_i^{(1,0,0)}(s^2) =& \left. \frac{\partial A_i(q^2,r^2,\theta) }{\partial q^2}\right\vert_{q^2 = r^2 = s^2, \, \theta = 2\pi/3 } \,, \nonumber\\
A_i^{(0,1,0)}(s^2) =& \left. \frac{\partial A_i(q^2,r^2,\theta) }{\partial r^2}\right\vert_{q^2 = r^2 = s^2, \, \theta = 2\pi/3 } \,, \nonumber\\
A_i^{(0,0,1)}(s^2) =& \left. \frac{\partial A_i(q^2,r^2,\theta) }{\partial \theta}\right\vert_{q^2 = r^2 = s^2, \, \theta = 2\pi/3 } \,.
\label{parder}
\end{align}
Analogous relations, not reported here, hold for the asymmetric configuration.

Before turning to the full construction of $\overline{\Gamma}^{\,\rm sym}_1 (s^2)$, 
we focus on certain global aspects that it 
displays at low momenta, which may be obtained from the above expressions with a moderate amount of effort.

\subsection{\label{sec:lowlimit} The low-momentum limit}

In particular, \1eq{BCstuff} allows one to deduce the {\it exact}
functional form of $\overline{\Gamma}_1^{\,\rm sym} (s^2)$
in the limit $s^2\to 0$. Indeed, a preliminary one-loop dressed analysis reveals that, in that limit, 
the combination $(s^4/4) Y_1(s^2) - (s^2/2) Y_4(s^2)$ yields a {\it constant} term, to be denoted by $C_t$. Moreover, 
 $s^2H_3(s^2)$, $s^2A_3(s^2)$ and $s^2 A_4(s^2)$ vanish,
while $A_1(0) = 1$, by virtue of the Taylor theorem~\cite{Taylor:1971ff}.
Consequently, the {\it leading contribution} originates from the  
combination \mbox{$F(s^2) [J(s^2) + (s^2/2) J^{\prime}(s^2)]$}. 

Then, it is straightforward to establish from \1eq{Junq} that
\mbox{$\lim\limits_{s^2 \to 0} s^2 J'(s^2) = a$}.
Thus, the asymptotic form of $\overline{\Gamma}_1^{\,\rm sym} (s^2)$ is given by 
\be 
\overline{\Gamma}_1^{\,\rm sym} (s^2) \assymto{s^2 \to 0} F(0)\left[ a\ln\left( \frac{s^2}{\mu^2} \right) + b + \frac{a}{2} \right] = {\tilde a} \ln\left( \frac{s^2}{\mu^2} \right) + {\tilde b} \,,
\label{eq:LIR}
\ee
where we have set $C_t=0$. Then, using the fact that the saturation value of the ghost dressing function is 
\mbox{$F(0) = 2.92$} when
one renormalizes at \mbox{$\mu = 4.3$ GeV}, together with the values for $a$ and $b$
quoted below \1eq{consabcd}, one finds ${\tilde a} = 0.303$ and ${\tilde b} = 2.87$.

\begin{figure}[t]
\includegraphics[width=0.5\textwidth]{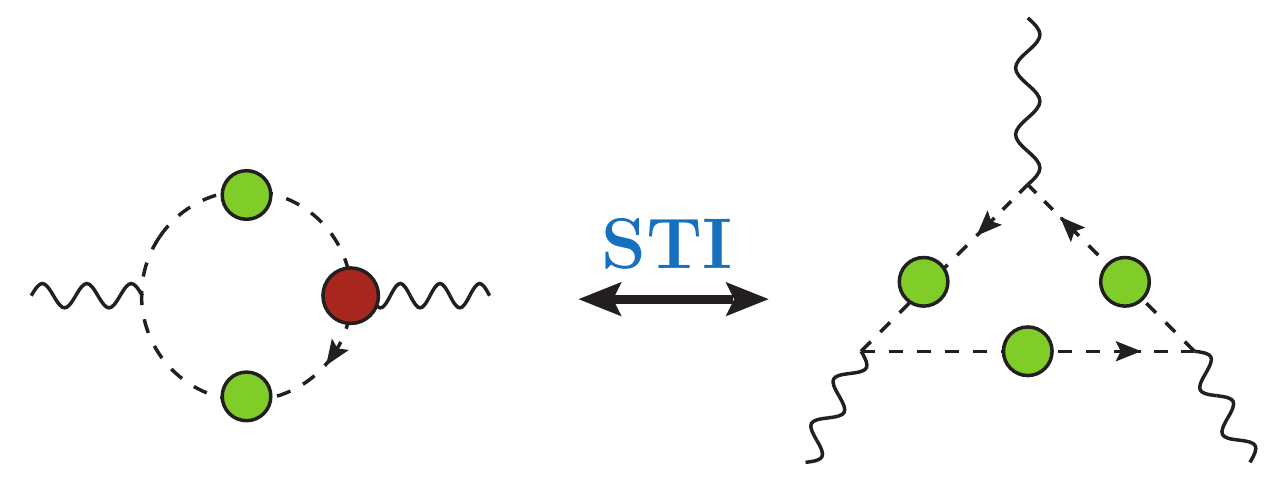} 
\caption{The ghost loop diagram contributing to the kinetic term $J(q^2)$, and the
ghost triangle diagram entering in the skeleton expansion of three-gluon vertex. Both 
diagrams are connected by the STI of \1eq{stig},
which imposes the equality of the corresponding unprotected logarithms.}
\label{fig:sti}
\end{figure}

In the asymmetric case, a similar procedure may be employed to fully determine the behavior of $\overline{\Gamma}_3^{\,\rm asym} (q^2)$ for small $q^2$, leading to
\be 
\overline{\Gamma}_3^{\,\rm asym} (q^2) \assymto{q^2 \to 0} F(0)\left[ J(q^2) + \lim_{q^2\to 0} q^2 J'(q^2) \right] = \tilde{a}\ln\left(\frac{q^2}{\mu^2}\right) + (b + a)F(0) \,. 
\label{eq:aslog}
\ee

It is important to clarify at this point that, in a  
{\it bona fide} SDE analysis of the three-gluon vertex~\cite{Schleifenbaum:2004id,Huber:2012kd,Aguilar:2013xqa,Huber:2012zj,Blum:2014gna,Eichmann:2014xya,Williams:2015cvx}, the asymptotic behavior found in \1eq{eq:LIR}
emerges from the ghost triangle diagram, shown in Fig.~\ref{fig:sti}, which furnishes an unprotected logarithm. 
Of course, in the BC construction followed in~\cite{Aguilar:2019jsj} and here, no vertex diagrams are considered; instead, 
the corresponding unprotected logarithm originates from the ghost loop diagram contributing to $J(q^2)$, shown in Fig.~\ref{fig:sti}, which is related to the ghost triangle diagram by the STI of \1eq{stig}, as shown schematically in Fig.~\ref{fig:sti}.

Note that the logarithms appearing in both \1eq{eq:LIR} and \1eq{eq:aslog} are multiplied by the same coefficient, namely 
$\tilde{a}$; this is a direct consequence of the fact that, in the Landau gauge, 
the ghost-gluon scattering kernel, $H_{\nu\mu}$,
assumes its tree level value when the momentum of its ghost leg vanishes, in compliance with the well-known
Taylor theorem. In particular, the $A_i$ enter into the BC solution with  
various permutations of $(q,r,p)$ in their arguments. 
Since in both cases considered all momenta eventually vanish,
the substitution \mbox{$A_1 \to 1$} and \mbox{$t^2A_{i}\to 0$} with \mbox{$i = 2,3,4,5$} is eventually triggered,
where $t$ denotes any of these momenta \footnote{The equality of the leading logarithms holds also  
perturbatively; however, in general,  
the $A_i$  cannot be set {\it individually} to their tree-level values, due to their higher rate of IR divergence. 
Nonperturbatively, the presence of a gluon mass scale attenuates these divergences~\cite{Aguilar:2018csq}, thus validating these substitutions.}. Specifically, one gets 
\begin{align}
X_1(s^2) &\assymto{s^2 \to 0} F(s^2) J(s^2) \,, \qquad &X_3(s^2) \assymto{s^2 \to 0} - F(s^2) J'(s^2) \,, \nonumber \\
X_1(q^2,q^2,\pi) &\assymto{q^2 \to 0} F(q^2) J(q^2) \,, \qquad &X_3(q^2,q^2,\pi) \assymto{q^2 \to 0} - F(0) J'(q^2) \,,
\end{align}
and the results of \2eqs{eq:LIR}{eq:aslog} follow straightforwardly.

Note that, within a self-consistent renormalization scheme, the coefficient ${\tilde a}$ 
multiplying the IR divergent logarithm is common to both $\overline{\Gamma}_3^{\,\rm asym} (q^2)$ and 
$\overline{\Gamma}_3^{\,\rm asym} (q^2)$. However, the conditions 
\mbox{$\overline{\Gamma}^{\, \rm sym}_1(\mu^2) = \overline{\Gamma}^{\, \rm asym}_3(\mu^2) = 1$},
enforced on the lattice data, cannot be simultaneously accommodated within a single scheme. Thus, 
the corresponding ${\tilde a}$ differ by a {\it finite} renormalization constant, which deviates very slightly from unity.

Let us finally point out that the qualitative analysis presented in this subsection 
remains valid even when $C_t\neq 0$, except for
the location of the zero crossing, which will be shifted in a direction and by an amount 
that depend on the sign and size of this constant.

\subsection{\label{sec:comp} Comparison with the lattice and further discussion}

\begin{figure}[t]
\centering 
\includegraphics[scale=0.27]{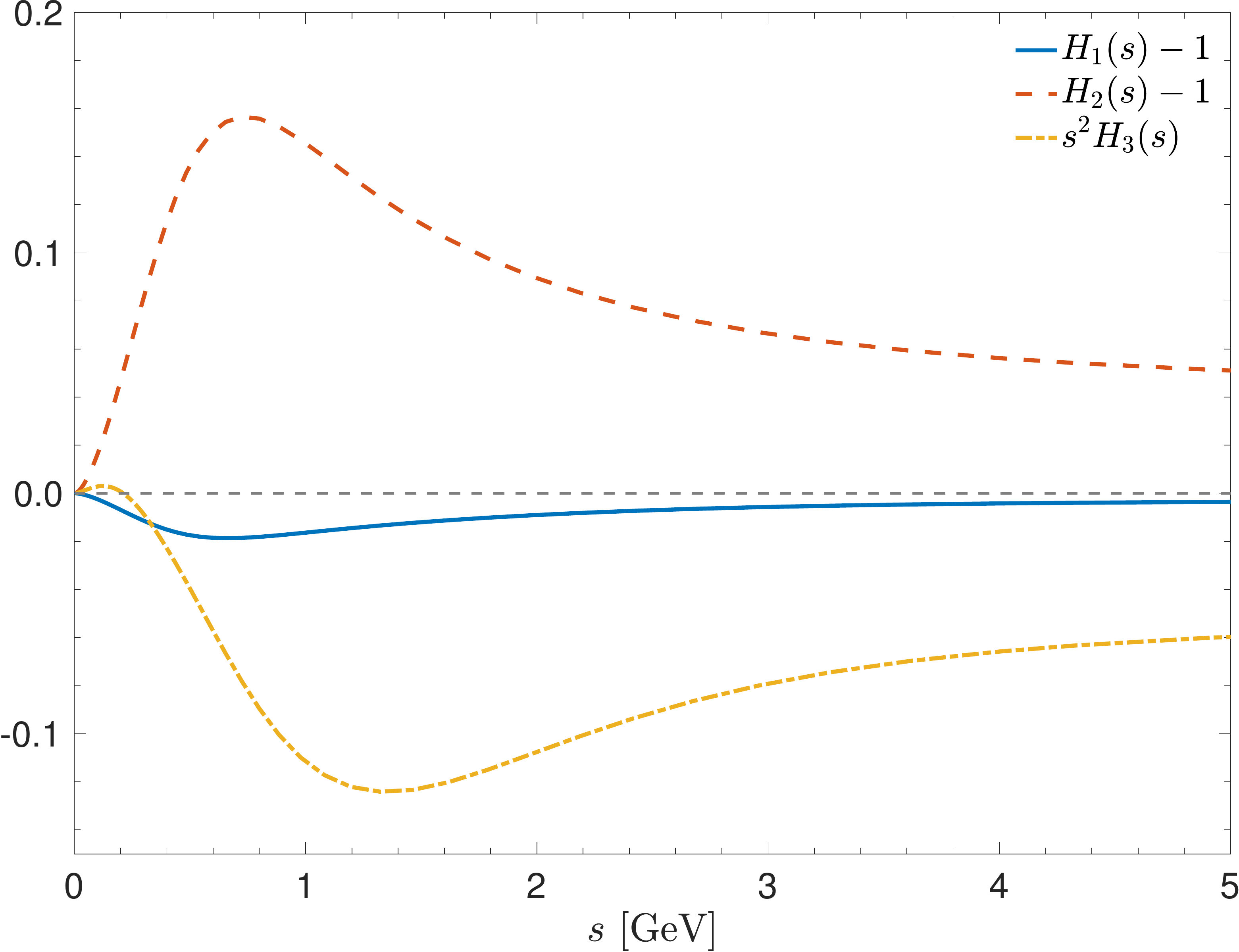}
\caption{The deviations of the combinations $H_1(s^2)$, $H_2(s^2)$, and $s^2H_3(s^2)$, defined in the Eq.~\eqref{theH}, from their tree-level counterparts.}
\label{fig:Hi}
\end{figure}

Next, we proceed to the full determination of $\overline{\Gamma}_1^{\,\rm sym} (s^2)$ and $\overline{\Gamma}_3^{\,\rm asym}(q^2)$
from the set of formulas given above [in particular Eqs.~(\ref{eq:GammaSym_Xi})~and~(\ref{eq:asyGamma}), together with \3eqs{BCstuff}{theH}{parder}]. In order to accomplish this task,
the functions $H_1(s^2)$, $H_2(s^2)$, and $s^2H_3(s^2)$ must be computed from their defining equations, 
given in \1eq{theH}. This, in turn, requires the determination of the
form factors $A_1(s^2)$, $A_3(s^2)$, and $A_4(s^2)$, and the corresponding derivatives;
since the impact of the unquenching effects
on the ghost sector is expected to be rather small~\cite{Ayala:2012pb}, for simplicity we use the
quenched $A_i$ of~\cite{Aguilar:2018csq}.
The final $H_1(s^2)$, $H_2(s^2)$, and $s^2H_3(s^2)$ are shown in  Fig.~\ref{fig:Hi}, for the symmetric configuration; 
similar results have been obtained for the asymmetric case (not shown). 
 
\begin{figure}[t]
\begin{minipage}[b]{0.45\linewidth}
\centering 
\hspace{-1.5cm}
\includegraphics[scale=0.27]{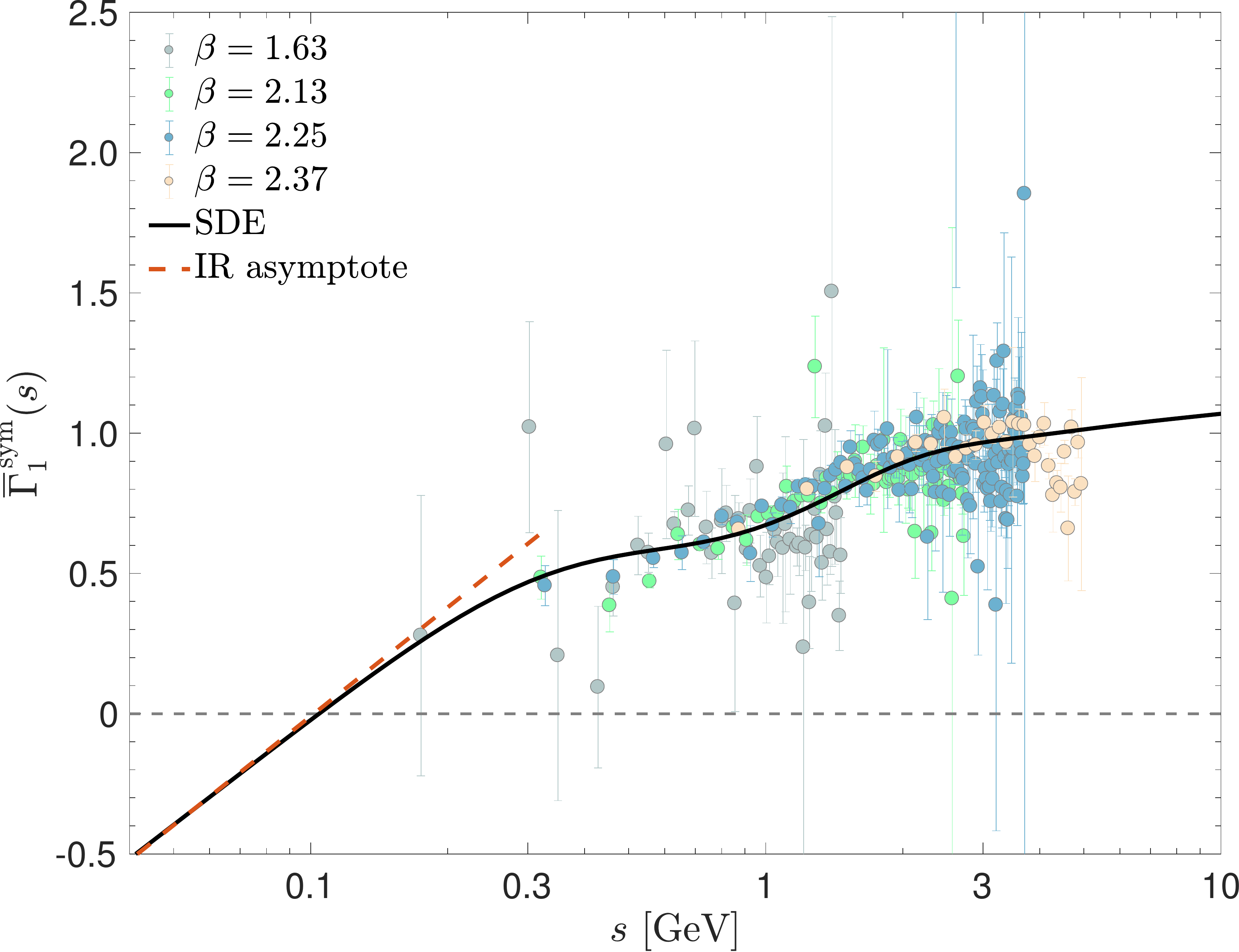}
\end{minipage}
\hspace{0.15cm}
\begin{minipage}[b]{0.45\linewidth}
\includegraphics[scale=0.27]{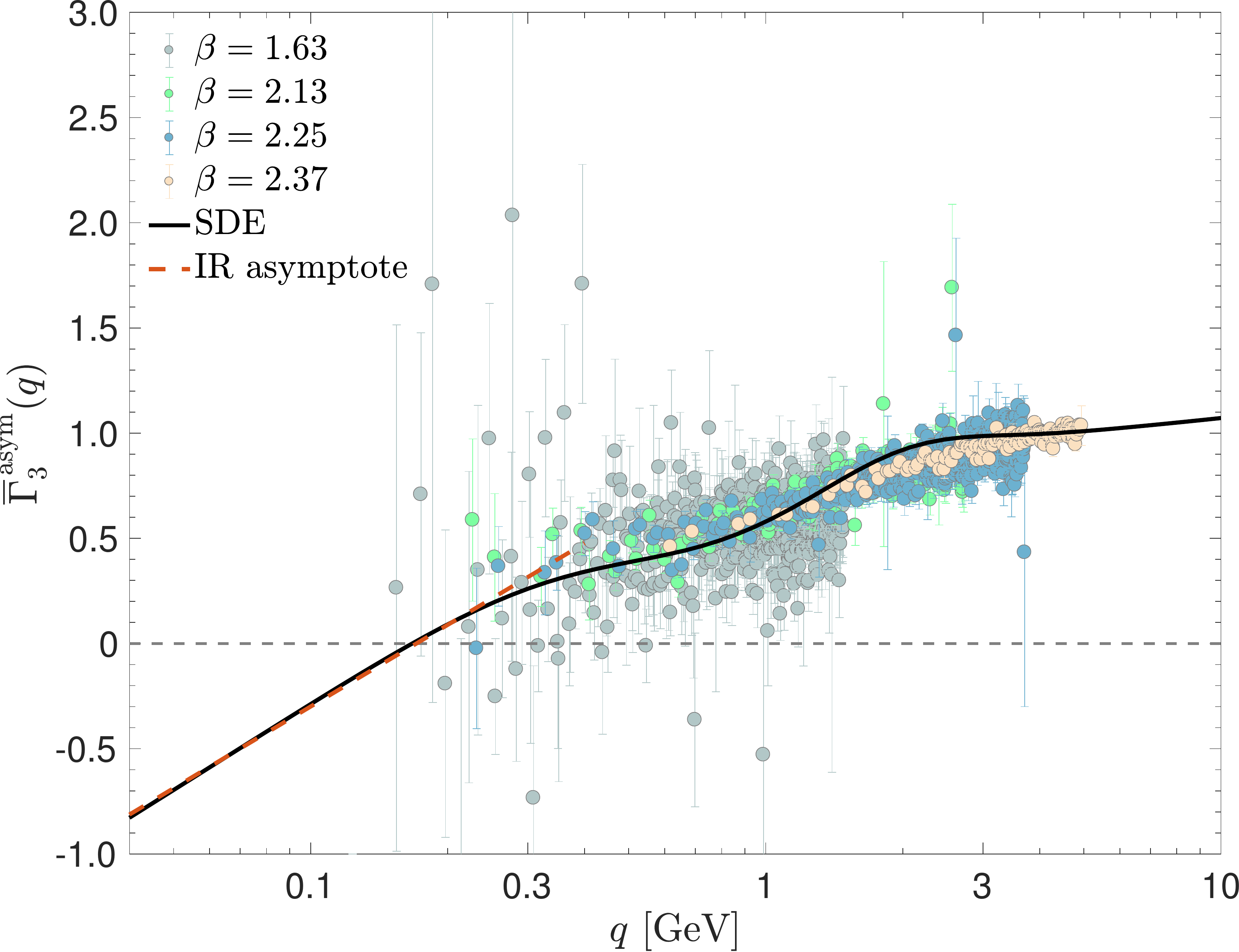}
\end{minipage}
\caption{ Left panel: The form factor of the three-gluon vertex in 
the symmetric configuration, \mbox{$\overline{\Gamma}_1^{\,\rm sym} (s)$}, obtained from lattice QCD (circles in different colors for each setup) and  from the SDE-based  approach (black continuous curve). Right panel: The same for the asymmetric form factor 
\mbox{$\overline{\Gamma}_3^{\,\rm asym} (q)$}. The IR asymptotes (red dashed lines) are given by Eqs.~\eqref{eq:LIR} and~\eqref{eq:aslog}, respectively.}
\label{fig:1PI}
\end{figure}

The comparison between the final SDE-based prediction
for $\overline{\Gamma}_1^{\,\rm sym} (s^2)$ and $\overline{\Gamma}_3^{\,\rm asym}(q^2)$
and the corresponding  unquenched lattice data is shown in Fig.~\ref{fig:1PI}; we observe 
a very good agreement for the entire range of momenta. 
It is rather evident that the particular shape of 
$J(q^2)$, shown in the top right panel of  Fig.~\ref{fig:JMDelta} and given by Eq.~\eqref{Junq}, is largely responsible for the 
most characteristic features of the vertex form factor at intermediate and low momenta, namely its 
overall suppression with respect to its tree-level value, and the inevitable
(albeit hard to observe) reversal of sign (zero crossing) in the deep IR.

It is clear that, due to the well-known ambiguities related with the
scale setting~\cite{Sommer:1993ce,Capitani:1998mq,Becirevic:1998jp,Boucaud:2000ey}, 
direct comparisons between quenched and unquenched data may be quantitatively subtle.
Notwithstanding this caveat, the inclusion of quarks seems to moderate the
amount of suppression with respect to~\cite{Athenodorou:2016oyh}.
Specifically, the decrease observed between the renormalization point of $\mu = 4.3$ GeV 
[where \mbox{$\overline{\Gamma}^{\, \rm sym}_1(\mu^2) = \overline{\Gamma}^{\, \rm asym}_3(\mu^2) = 1$}]
and a typical IR momentum, say  \mbox{$q_{\s{I\!R}}$=300  MeV}, is given by 
\mbox{$\overline{\Gamma}_1^{\rm sym}(q_{\s{I\!R}}^2)$=0.47}       and
\mbox{$\overline{\Gamma}_1^{\rm asym}(q_{\s{I\!R}}^2)$=0.26}, to be compared with 
\mbox{$\overline{\Gamma}_1^{\rm sym}(q_{\s{I\!R}}^2)$=0.33}           and
\mbox{$\overline{\Gamma}_1^{\rm asym}(q_{\s{I\!R}}^2)$=0.2} for the  quenched case; thus, 
the observed suppression is reduced by about \mbox{25\%}. 
In addition, as expected from the corresponding displacement of $q_0$ at the level of the $J(q^2)$ 
[see Sec.~\ref{subsec:asymp}, point {\it (i)}],  
the zero crossing of both vertex configurations occurs at momenta lower compared to the
quenched case, in qualitative agreement with the analysis of~\cite{Williams:2015cvx}.
In particular, we find that the zero crossing moves from about \mbox{150 MeV} down to \mbox{105 MeV} (symmetric case) and from roughly \mbox{240 MeV} to about \mbox{170 MeV} (asymmetric case).


\begin{figure}[t]
\begin{minipage}[b]{0.45\linewidth}
\hspace{-1.5cm}
\centering
\includegraphics[scale=0.27]{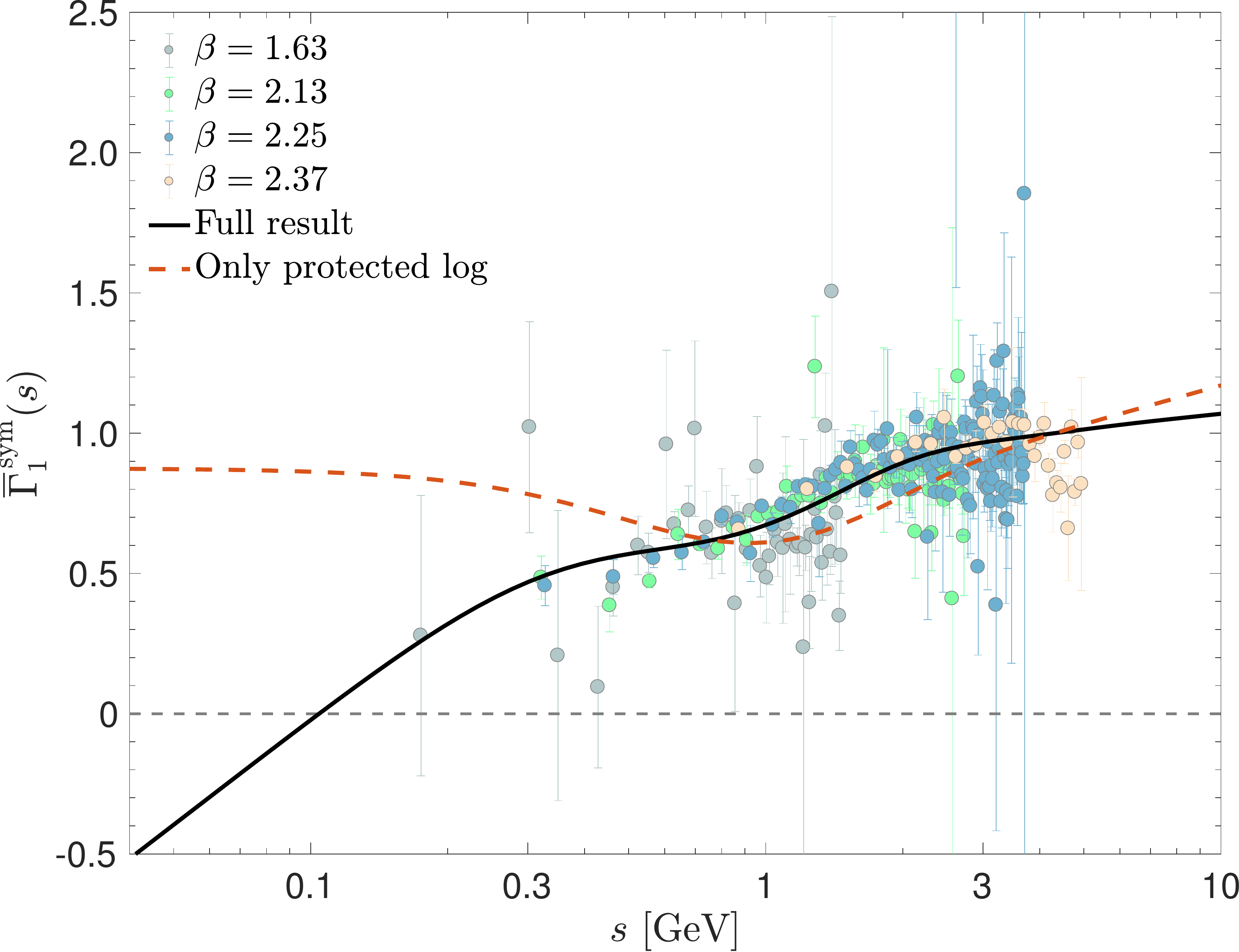}
\end{minipage}
\hspace{0.15cm}
\begin{minipage}[b]{0.45\linewidth}
\includegraphics[scale=0.27]{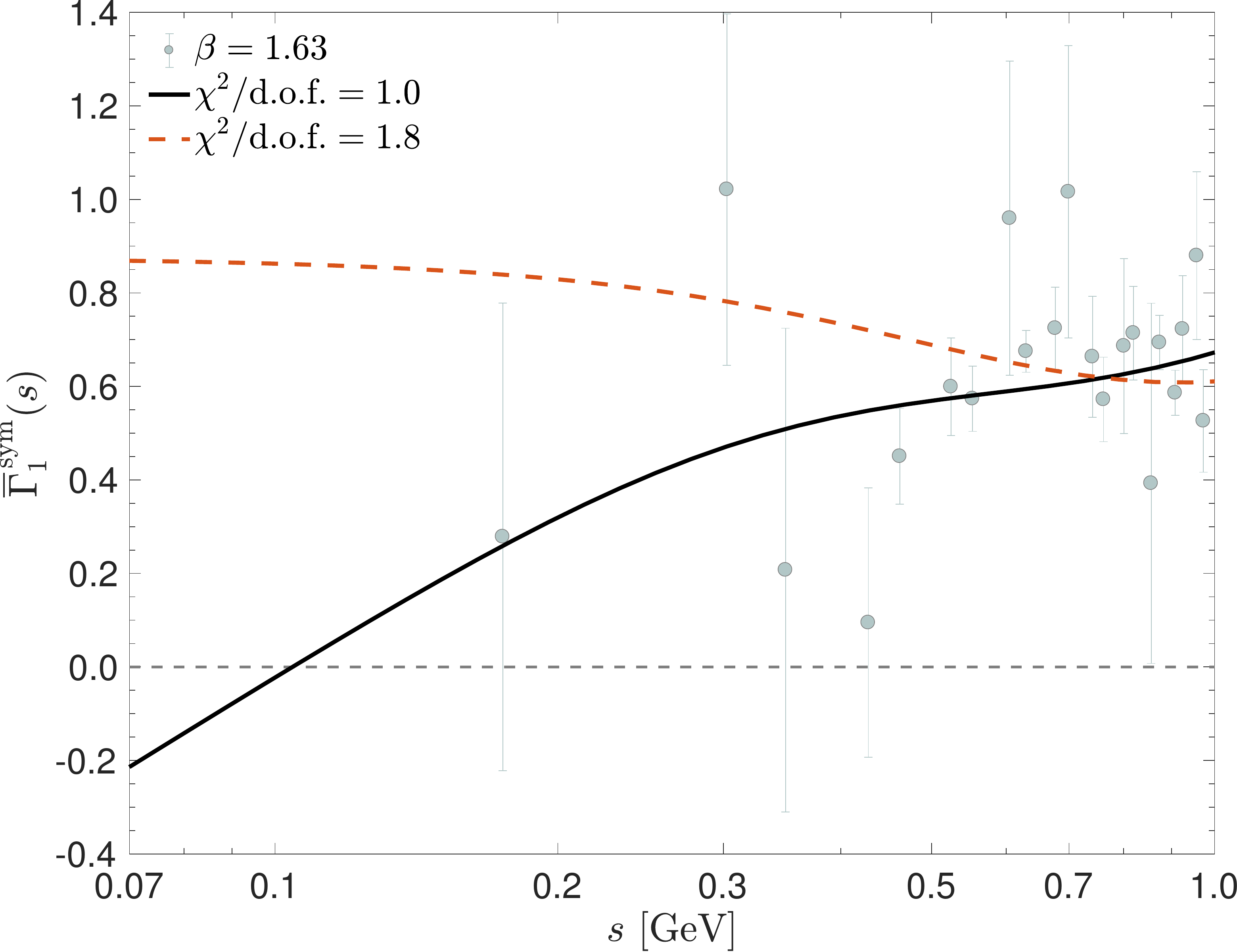}
\end{minipage}
\begin{minipage}[b]{0.45\linewidth}
\hspace{-1.5cm}
\includegraphics[scale=0.27]{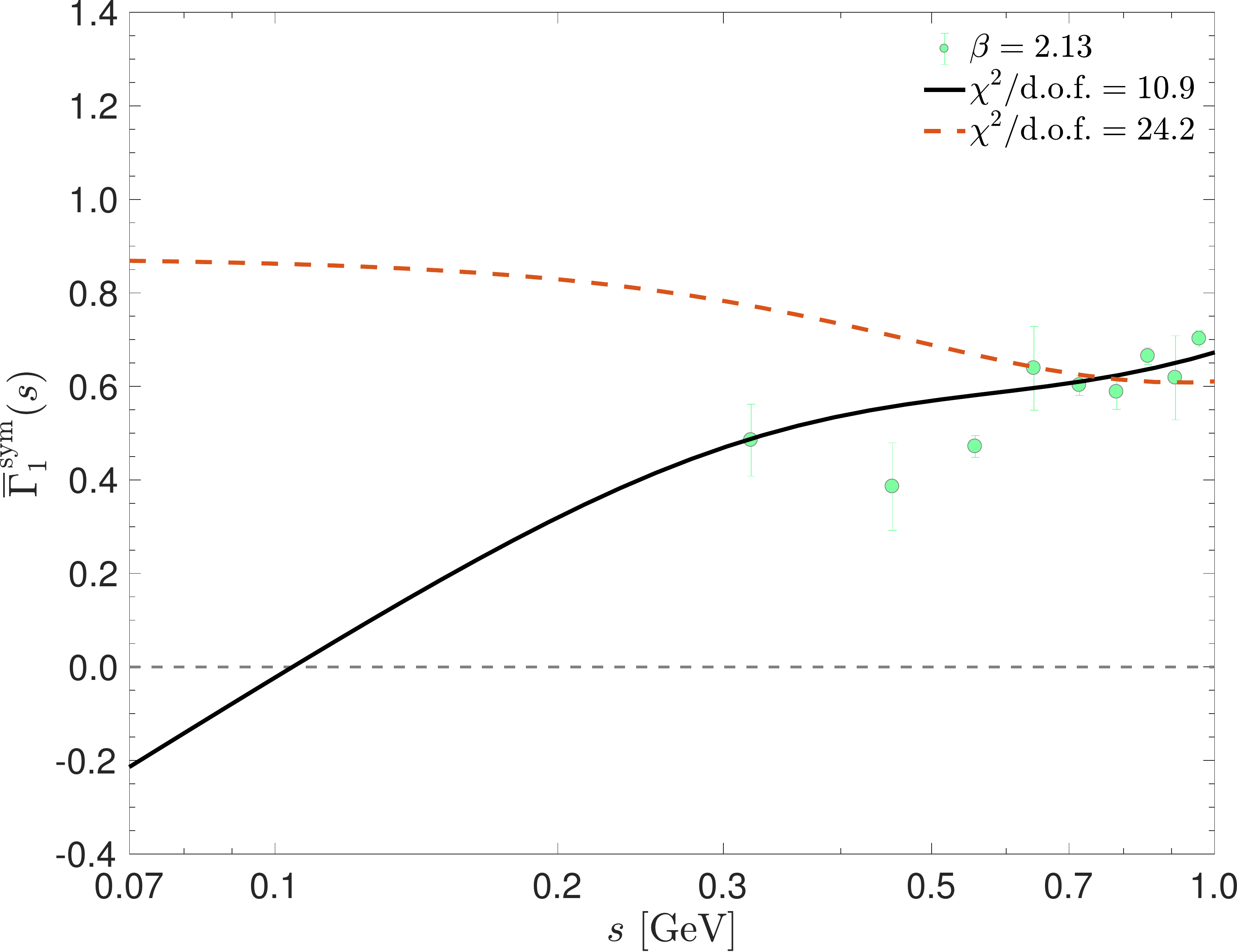}
\end{minipage}
\begin{minipage}[b]{0.45\linewidth}
\includegraphics[scale=0.27]{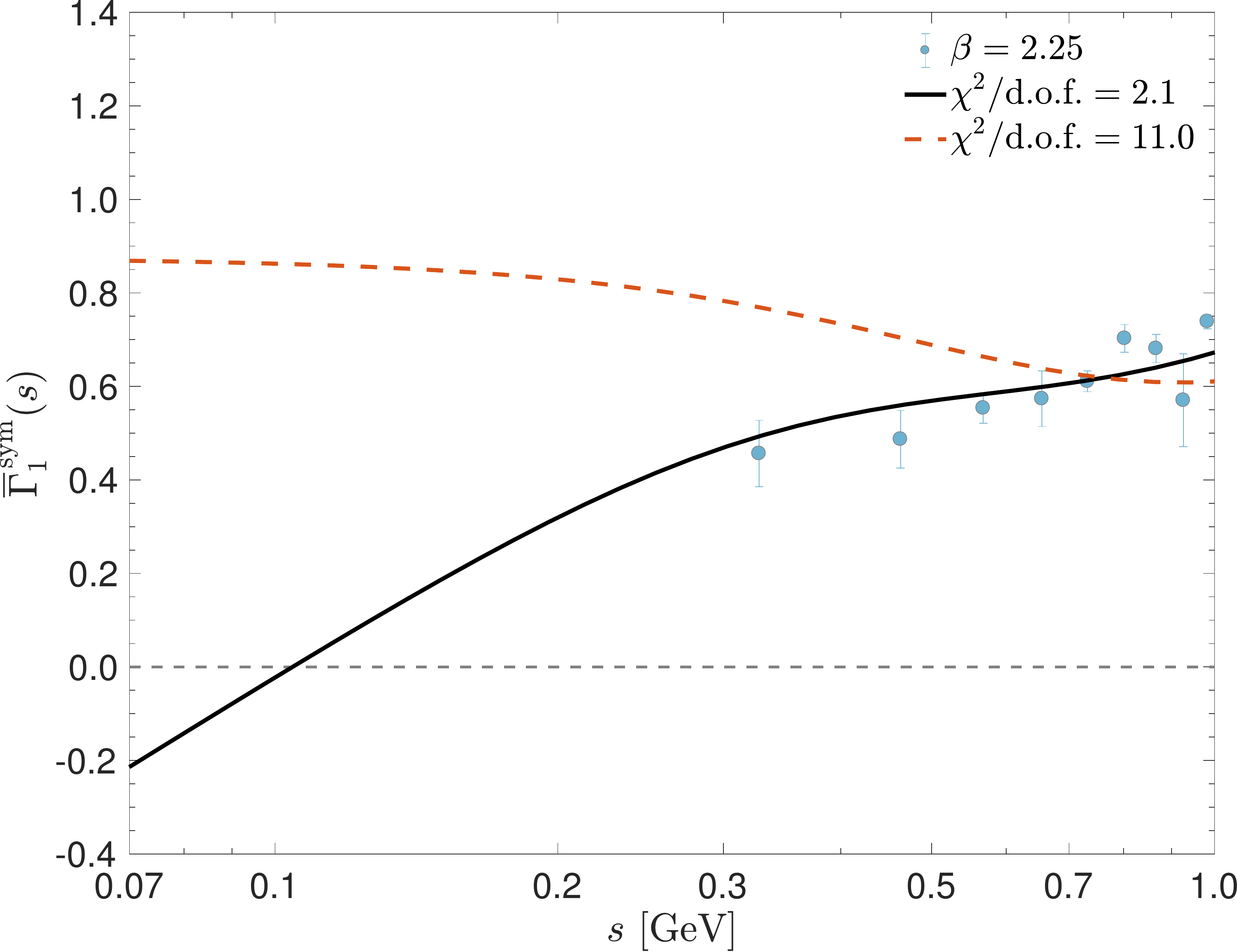}
\end{minipage}
\caption{ Top left plot: The effect of the unprotected logarithm of $J(q^2)$ [see Eq.~\eqref{Junq}] on  the IR behavior of $\overline{\Gamma}_1^{\,\rm sym} (s)$. The black continuous line is  
 $\overline{\Gamma}_1^{\,\rm sym} (s)$ derived using the full expression for $J(q^2)$, given by 
Eq.~\eqref{Junq}. The red dashed curve represents the case where the massless logarithm, appearing in Eq.~\eqref{Junq}, is neglected. Top right plot: The  \mbox{$\chi^2/$d.o.f.} when we compare the lattice data for $\beta=1.63$ with the results for $\overline{\Gamma}_1^{\,\rm sym} (s)$ with (black continuous)  and without (red dashed) the unprotected logarithm. Bottom left plot: The same as the previous panel but for the lattice data with $\beta=2.13$. Bottom right plot:  Same analysis, using the lattice data with $\beta=2.25$.}
\label{fig:protect}
\end{figure}

We next study in more detail the impact of 
the unprotected logarithm of $J(q^2)$  on the IR behavior of the vertex.
In particular, $\overline{\Gamma}_1^{\,\rm sym} (s^2)$ is computed by plugging into \1eq{BCstuff}
{\it (i)}  the {\it full} $J(q^2)$ given in Eq.~\eqref{Junq} (black continuous curve), and  
{\it (ii)}  a $J(q^2)$ {\it without} 
the term  \mbox{$(1/6)\ln\left(q^2/\mu^2\right)$} (red dashed curve).
As we can see in the top left panel of Fig.~\ref{fig:protect}, for momenta 
below \mbox{$800$ MeV} the unprotected logarithm starts
to dominate the behavior of $\overline{\Gamma}_1^{\,\rm sym} (s^2)$,
forcing not only its suppression but also its IR divergence.
In the remaining panels of Fig.~\ref{fig:protect} we show that the three sets of lattice data considered 
exhibit {\it individually} a  clear preference for  case {\it (i)}; in fact,
even in the least favorable case (top right panel), where the data are rather sparse and with sizable errors, 
the \mbox{$\chi^2/$d.o.f.} is  $1.8$ times smaller than that of case {\it (ii)}.

Finally, turning to the transverse part of $\Gamma_{\alpha\mu\nu}$,
it is clear that the corresponding form factors 
ought to be determined from a detailed SDE study, which is still pending.
In fact, the good coincidence found between the SDE-based 
prediction [with $Y_1(s^2)=Y_4(s^2)=0$] and the lattice must be interpreted with caution, 
especially in view of the sizable errors assigned to the data. Indeed, given the present precision,  
one may easily envisage how reasonably sized transverse contributions could be rather comfortably accommodated, 
provided they follow the general trend of the data.
We hope to report progress in this direction in the near future. 

\subsection{\label{sec:effcoup} Effective couplings}

It is rather instructive to study how the suppression of the three-gluon vertex manifests itself 
at the level of a typical {\it renormalization-group invariant} quantity, which is traditionally used to quantify 
the effective strength of a given interaction.

\begin{figure}[t]
\begin{minipage}[b]{0.45\linewidth}
\centering 
\hspace{-1.5cm}
\includegraphics[scale=0.27]{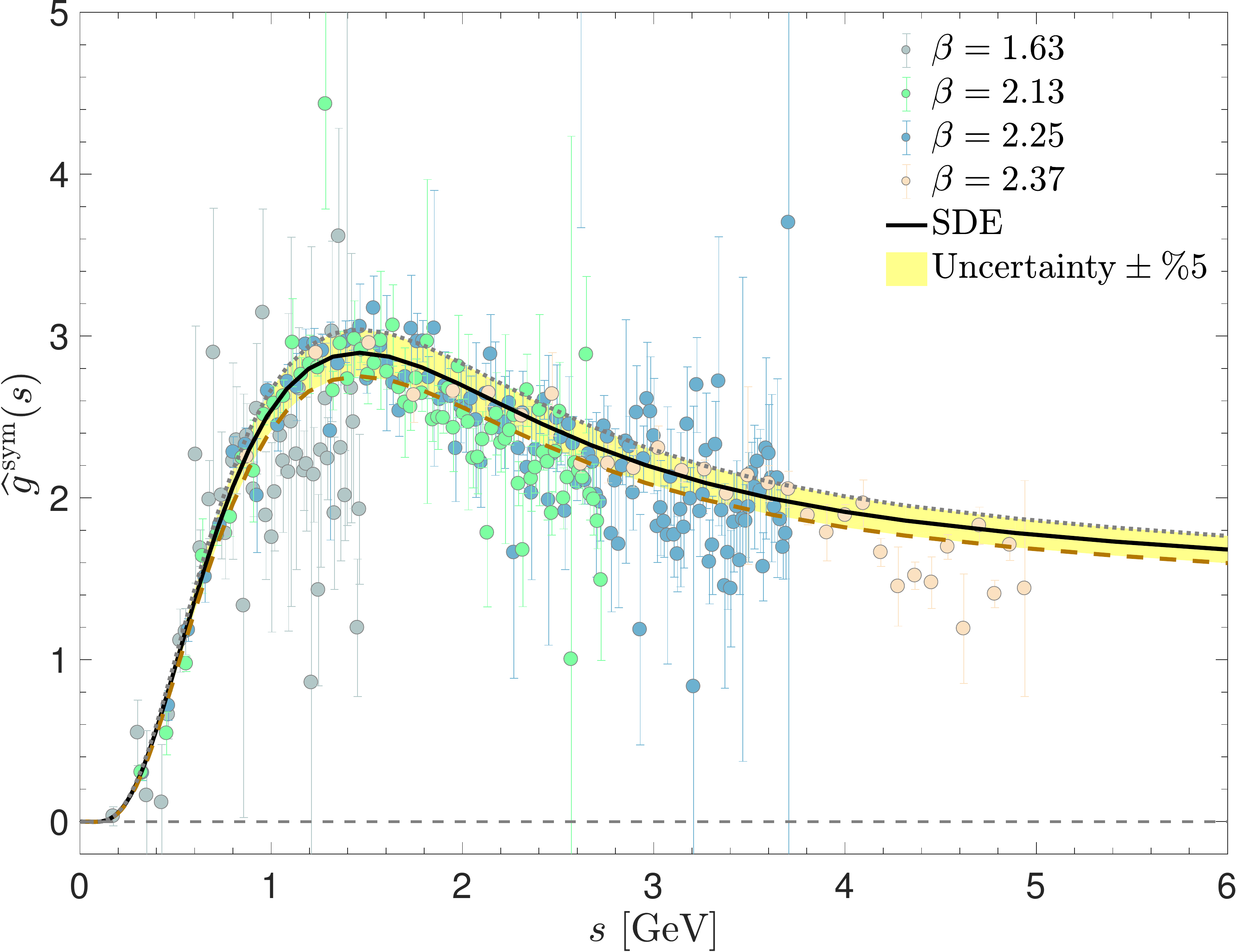}
\end{minipage}
\hspace{0.15cm}
\begin{minipage}[b]{0.45\linewidth}
\includegraphics[scale=0.27]{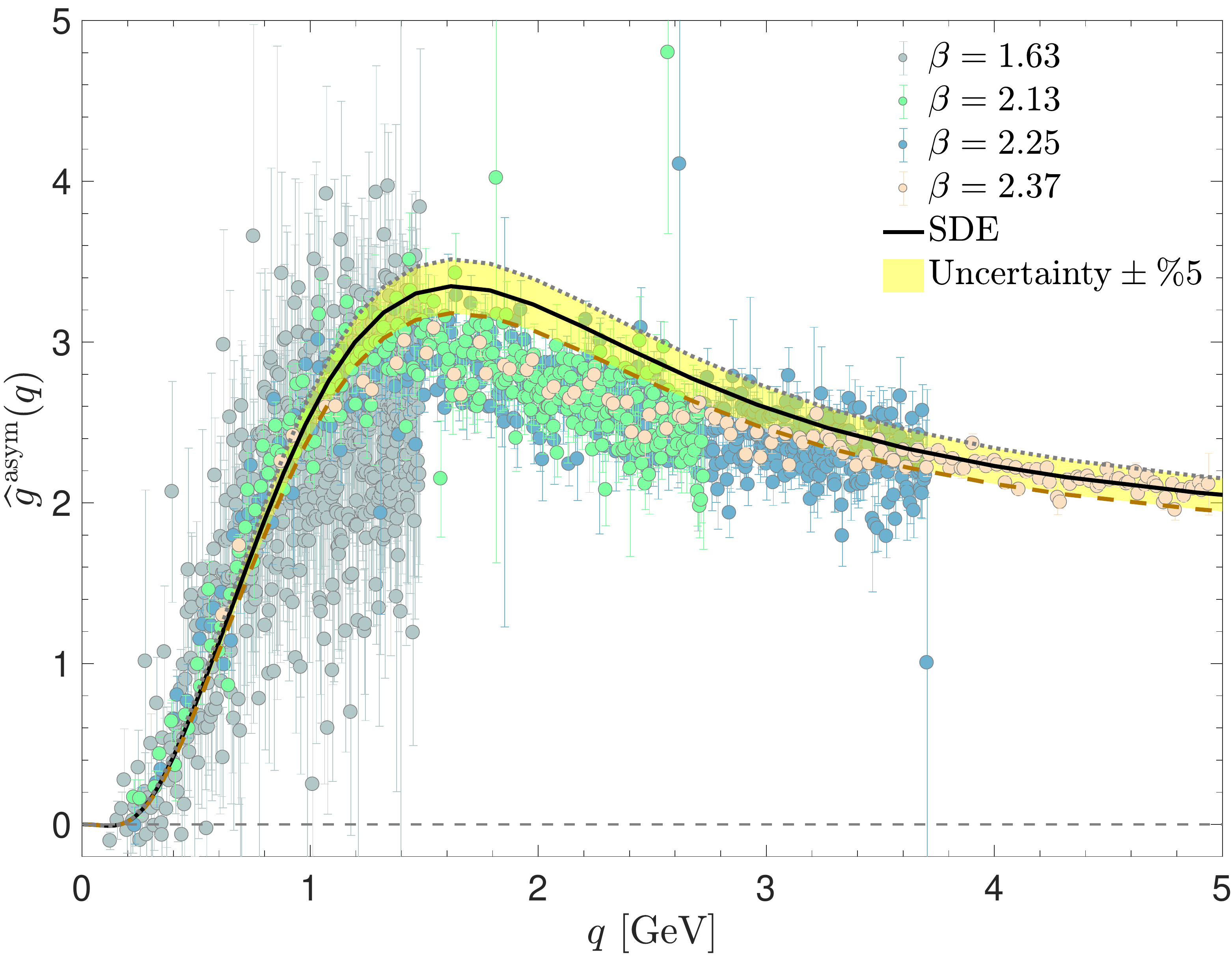}
\end{minipage}
\caption{ Left panel: The symmetric effective coupling, ${\widehat g}^{\,\rm sym}(s)$, defined in Eq.~\eqref{twocoupl}, obtained using the lattice data (full circles) and  the results of the SDE-based approach (black continuous line). The yellow band represents 
  how the SDE result for ${\widehat g}^{\,\rm sym}(s)$ changes when the value of $g$ at \mbox{$\mu$=4.3 GeV} has an uncertainty of $\pm 5\%$,
  with central values $g^{\,\rm sym}(\mu^2)=1.86$ [$\alpha^{\,\rm sym}(\mu^2) = 0.27$].
Right panel:  The same for
the asymmetric effective coupling, ${\widehat g}^{\,\rm asym}(q^2)$, with a central value $g^{\,\rm asym}(\mu^2)=2.16$ [$\alpha^{\,\rm asym}(\mu^2) = 0.37$].}
\label{fig:gs}
\end{figure}

To that end, we next consider the two effective couplings
related to $\overline{\Gamma}_1^{\,\rm sym} (s^2)$ and $\overline{\Gamma}_3^{\,\rm asym}(q^2)$,
to be denoted by ${\widehat g}^{\,\rm sym}(s^2)$ and ${\widehat g}^{\,\rm asym}(q^2)$, respectively.
In particular, following standard definitions~\cite{Athenodorou:2016oyh,Mitter:2014wpa,Fu:2019hdw}, we have 
\be
{\widehat g}^{\,\rm sym}(s^2) = g^{\,\rm sym}(\mu^2)\, s^3\,\overline{\Gamma}_1^{\,\rm sym} (s^2)\Delta^{3/2}(s^2)\,,
\qquad
{\widehat g}^{\,\rm asym}(q^2) = g^{\,\rm asym}(\mu^2)\, q^3 \,\overline{\Gamma}_3^{\,\rm asym}(q^2)\Delta^{3/2}(q^2)\,.
\label{twocoupl}
\ee
We emphasize that these two couplings may be recast in the form 
\be
{\widehat g}^{\,\rm sym}(s^2) =  s^3 \frac{T^{\,\rm sym}(s^2)}{\Delta^{3/2}(s^2)} \,, 
\qquad
{\widehat g}^{\,\rm asym}(q^2) =  q^3 \, \frac{T^{\,\rm asym}(q^2)}{\Delta(0) \Delta^{1/2}(q^2)} \,,
\ee 
thus making contact with the corresponding  definitions employed within the MOM schemes \mbox{\cite{Alles:1996ka,Boucaud:1998bq}}. Turning to their computation, we use for the ingredients entering in the above definitions
both lattice data as well as the corresponding SDE-derived quantities; the results obtained are displayed in Fig.~\ref{fig:gs}. 

\begin{figure}[t]
\begin{minipage}[b]{0.45\linewidth}
\hspace{-1.5cm}
\centering
\includegraphics[scale=0.27]{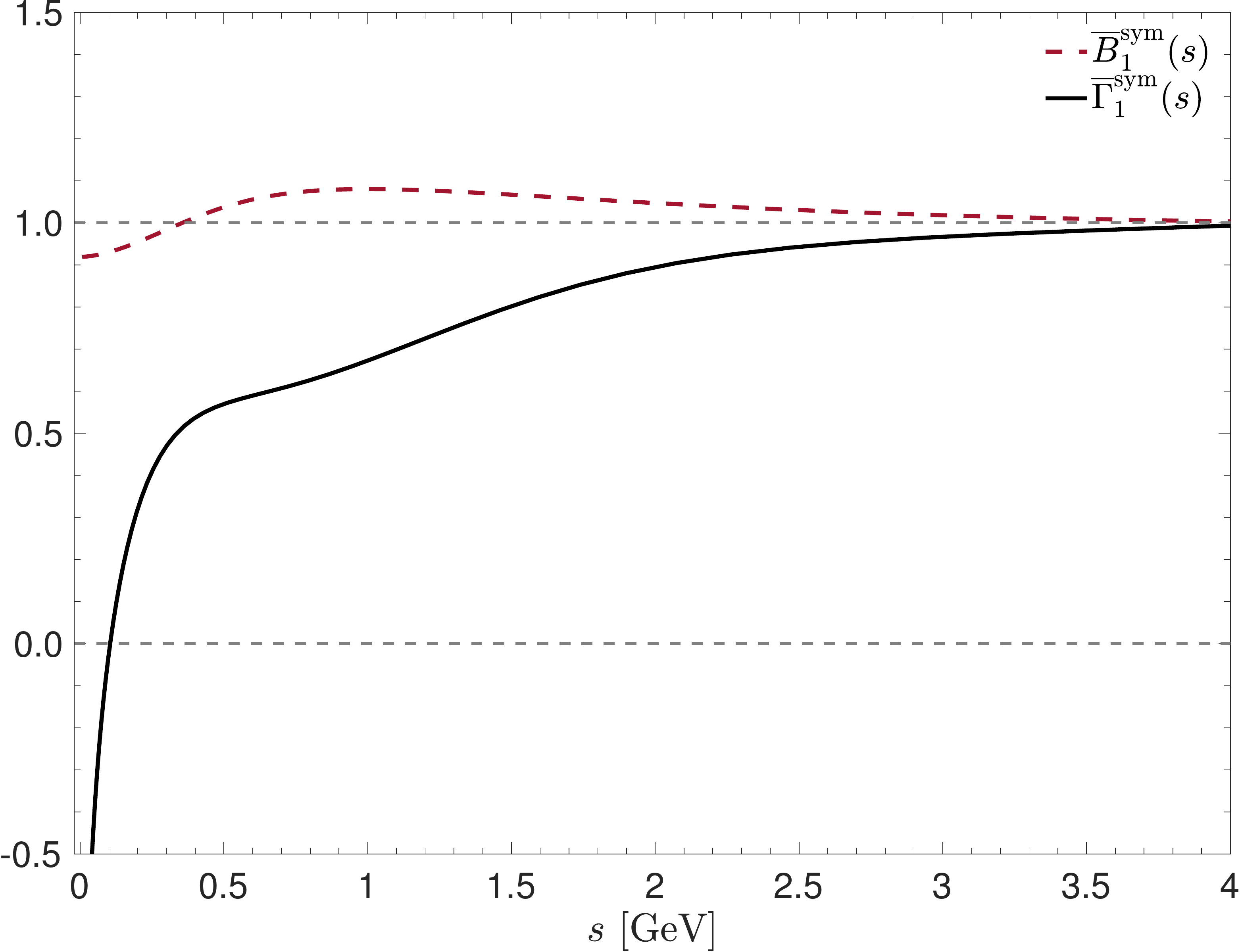}
\end{minipage}
\hspace{0.15cm}
\begin{minipage}[b]{0.45\linewidth}
\includegraphics[scale=0.27]{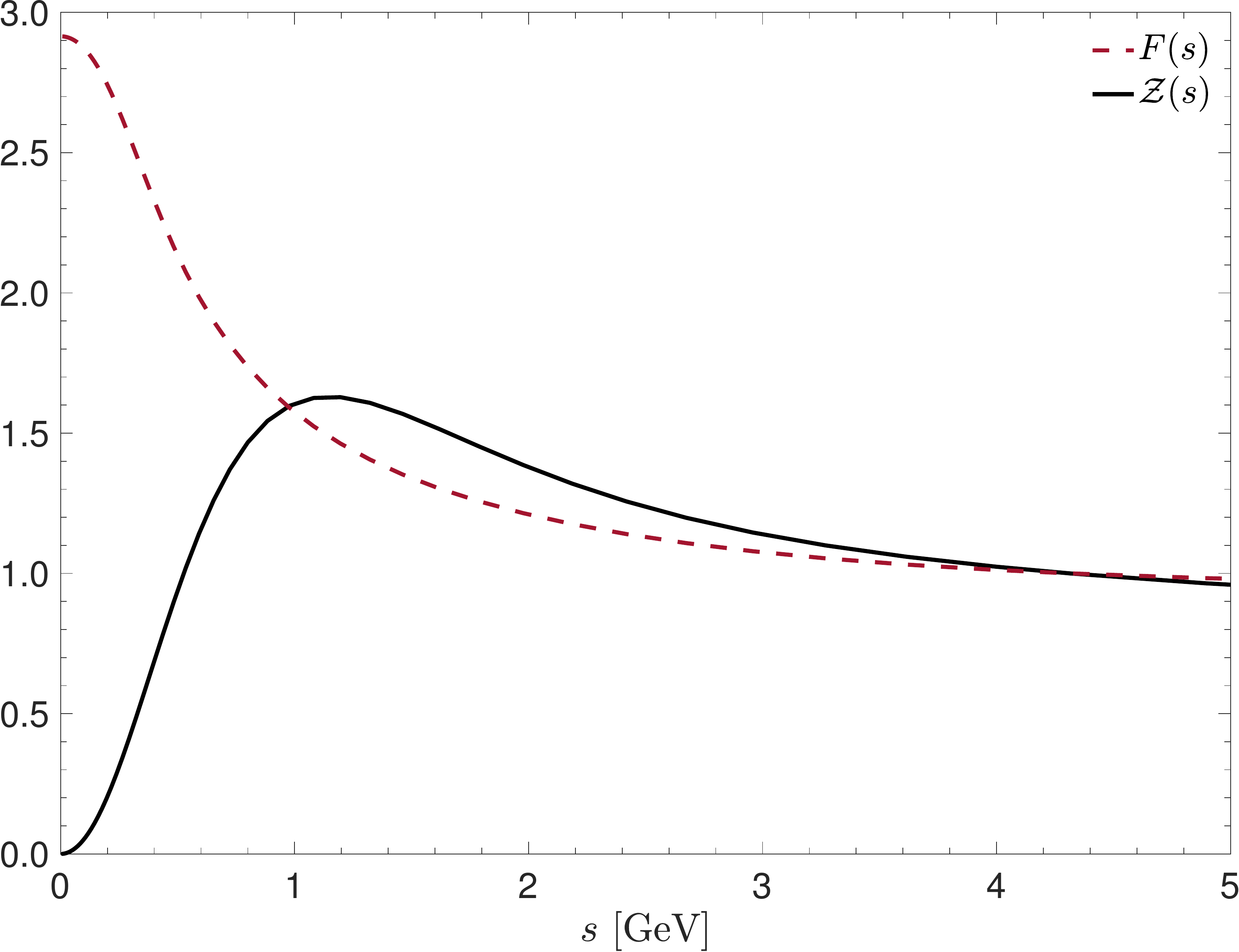}
\end{minipage}
\caption{Left panel: $\overline{\Gamma}_1^{\,\rm sym}(s)$ (black continuous line) compared with the form factor \mbox{$\overline{B}_1^{\,\rm sym} (s)$} of   
the ghost-gluon vertex (red dashed curve),  in the symmetric configuration.  Right panel: The gluon and ghost dressing functions, ${\cal Z}(s)$ and $F(s)$, respectively.}
\label{fig:form}
\end{figure}

\begin{figure}[t]
\begin{minipage}[b]{0.45\linewidth}
\hspace{-1.7cm}
\includegraphics[scale=0.27]{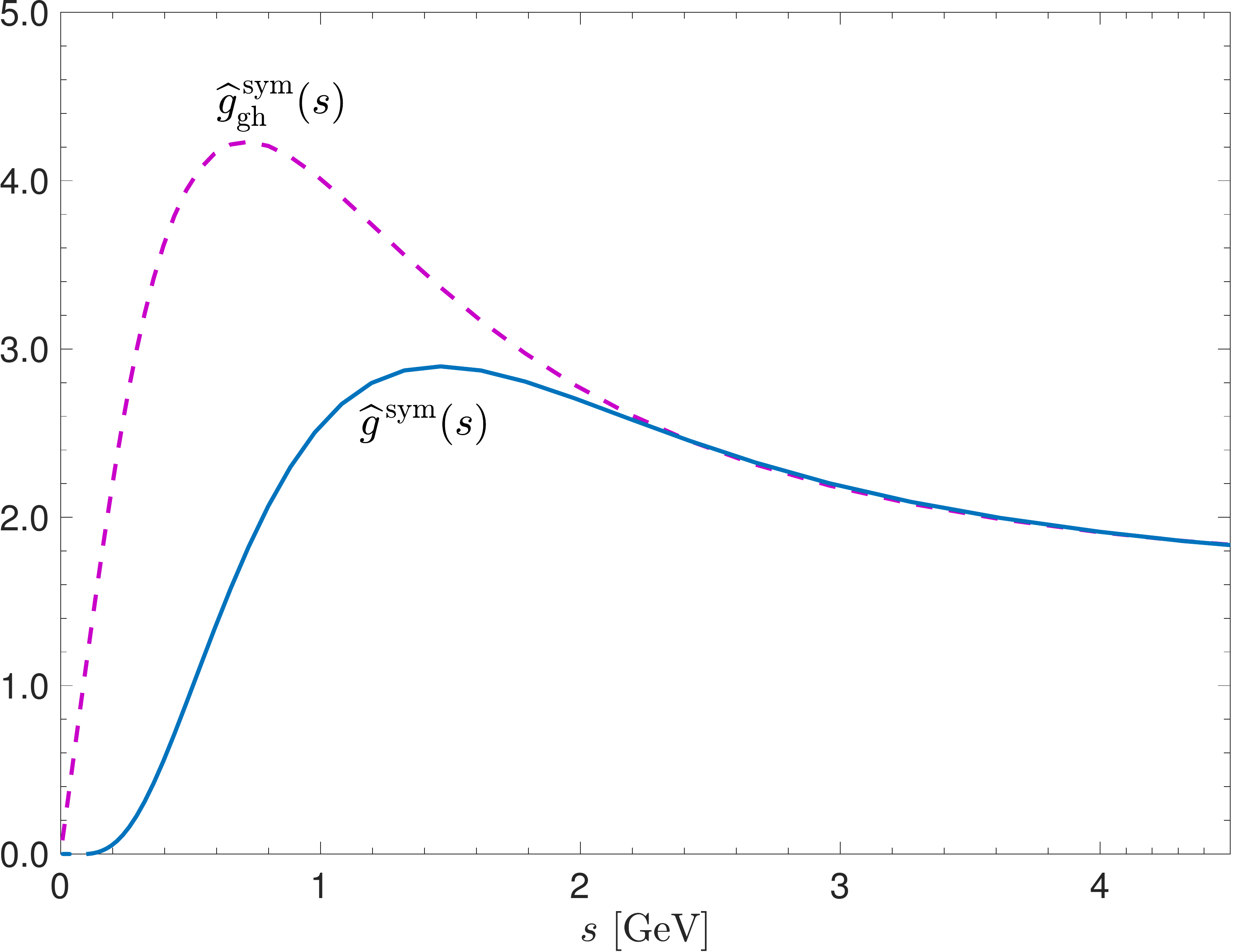}
\end{minipage}
\begin{minipage}[b]{0.45\linewidth}
\includegraphics[scale=0.27]{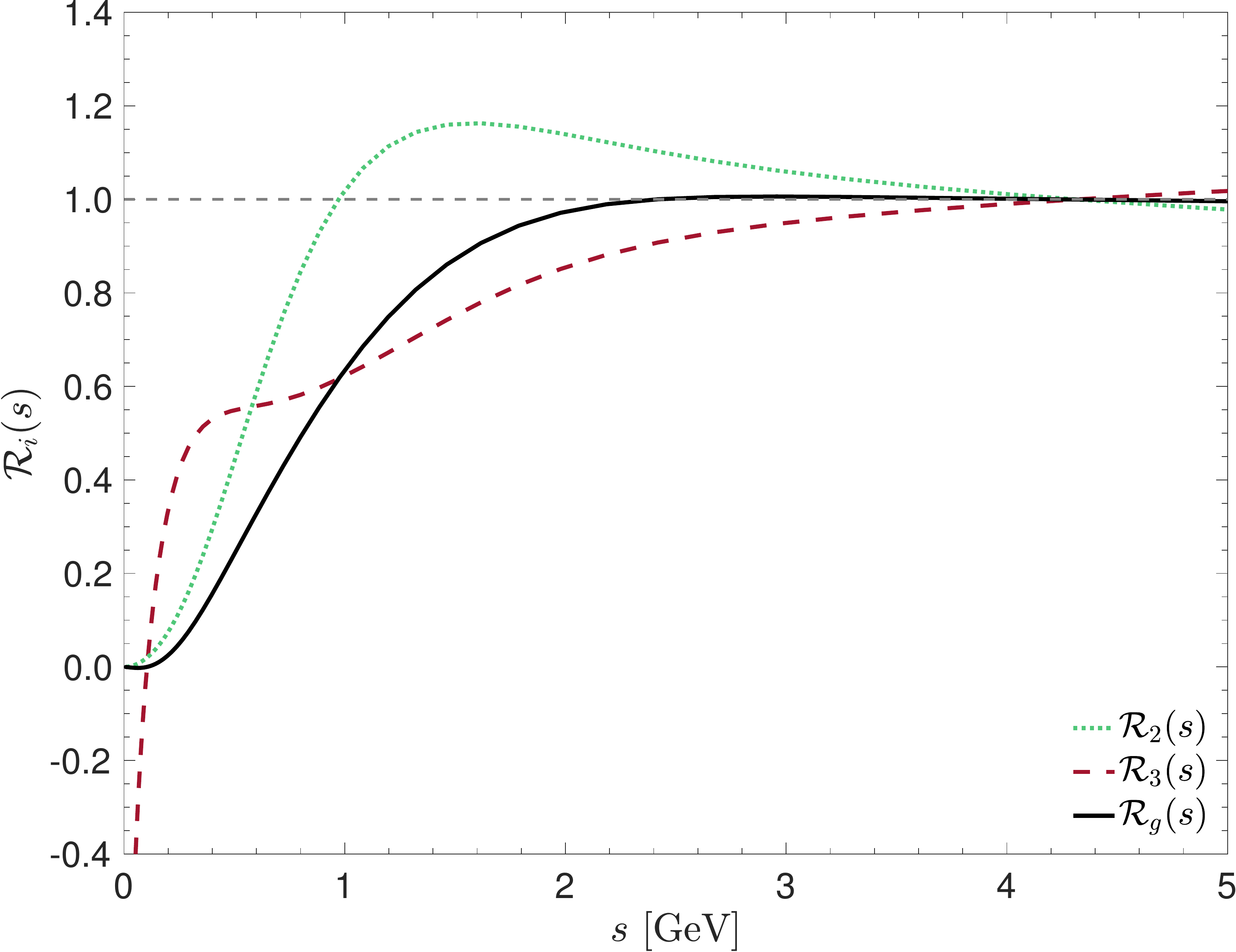}
\end{minipage}
\caption{Left panel: The comparison of the effective couplings, ${\widehat g}^{\,\rm sym}(s)$ (blue solid line) and ${\widehat g}_{\rm gh}^{\,\rm sym}(s)$ (magenta dashed), defined from the three-gluon vertex, Eq.~\eqref{twocoupl}, and from the ghost-gluon vertex,
Eq.~\eqref{ghcoupl}, respectively. Right panel: The ratios  ${\cal R}_g(s)$ (black continuous),  ${\cal R}_2(s)$ (green dotted), and ${\cal R}_3(s)$ (red dashed) introduced in the Eq.~\eqref{ratios}.}
\label{fig:coup_ratio}
\end{figure}

It is interesting to carry out a direct comparison of the effective coupling, ${\widehat g}^{\,\rm sym}(s^2)$, with the corresponding quantity, ${\widehat g}_{\rm gh}^{\,\rm sym}(s^2)$,
associated with the ghost-gluon vertex in the symmetric configuration. Specifically\footnote{Using the formulas of~\cite{Chetyrkin:2000fd}, one finds that $g^{\rm sym}(\mu^2)/g^{\rm sym}_{\rm gh}(\mu^2) =1.03$ at $\mu=4.3$ GeV, 
which justifies the use of $g^{\rm sym}(\mu^2)$ instead of $g^{\rm sym}_{\rm gh}(\mu^2)$ in the definition of \1eq{ghcoupl}. },    
\be
{\widehat g}_{\rm gh}^{\,\rm sym}(s^2) =  g^{\,\rm sym}(\mu^2)\, s \, \overline{B}_1^{\,\rm sym} (s^2) F(s^2) \Delta^{1/2}(s^2) \,, 
\label{ghcoupl}
\ee
where $\overline{B}_1^{\,\rm sym} (s^2)$ denotes the form factor proportional to the
tree-level  component of the ghost-gluon vertex, renormalized at the same MOM point,  
\mbox{$\mu = 4.3$ GeV}. The functional form used for $\overline{B}_1^{\,\rm sym} (s^2)$ 
has been obtained from the analysis of~\cite{Aguilar:2018csq} and it is shown in the
left panel of Fig.~\ref{fig:form}.   
The two couplings are displayed in the left panel of Fig.~\ref{fig:coup_ratio}; clearly,
as the momentum $s$ decreases, ${\widehat g}^{\,\rm sym}(s^2)$
becomes considerably smaller than  ${\widehat g}_{\rm gh}^{\,\rm sym}(s^2)$.

In order to analyze in detail the origin of this relative suppression,    
it is advantageous to introduce the gluon dressing function, 
${\cal Z}(q^2)$, defined as ${\cal Z}(q^2) = q^2 \Delta(q^2)$, which is shown on the right panel  of Fig.~\ref{fig:form}, together
with the corresponding quantity for the ghost propagator, $F(q^2)$, introduced in \1eq{eq:ghost_dressing}. 
Then, the two effective couplings assume the form
\be
{\widehat g}^{\,\rm sym}(s^2) = g^{\,\rm sym}(\mu^2)\, \overline{\Gamma}_1^{\,\rm sym} (s^2) {\cal Z}^{3/2}(s^2)\,,
\qquad
{\widehat g}_{\rm gh}^{\,\rm sym}(s^2)= g^{\,\rm sym}(\mu^2)\, \overline{B}_1^{\,\rm sym} (s^2) F(s^2) {\cal Z}^{1/2}(s^2)\,.
\label{coupwithZ}
\ee
We next consider the ratio of these two couplings,
\be
{\cal R}_g(s^2) = {\widehat g}^{\,\rm sym}(s^2)/{\widehat g}_{\rm gh}^{\,\rm sym}(s^2) =
\underbrace{\left[{\cal Z}(s^2)/F(s^2)\right]}_{{\cal R}_2(s^2)}
\underbrace{\left[\overline{\Gamma}_1^{\,\rm sym} (s^2)/\overline{B}_1^{\,\rm sym} (s^2)\right]}_{{\cal R}_3(s^2)} \,,
\label{ratios}
\ee
where the partial ratios ${\cal R}_2(s^2)$ and ${\cal R}_3(s^2)$ quantify the relative contribution from the two- and three-point sectors, respectively,
at the various momentum scales involved.
The three ratios, ${\cal R}_g(s^2)$, ${\cal R}_2(s^2)$, and ${\cal R}_3(s^2)$ are shown on the right panel of Fig.~\ref{fig:coup_ratio}. 

Interestingly, ${\cal R}_2(s^2)$  and ${\cal R}_3(s^2)$ are smaller than $1$  for \mbox{$s<880$ MeV} and
\mbox{$s<4.3$ GeV}, respectively. Therefore,  in the region of momenta 
between \mbox{$(0-880)$ MeV}, the  suppression of ${\widehat g}^{\,\rm sym}(s^2)$ emerges as a combined effect of 
{\it both} the two- and the three-point sectors, whereas,  from $880\,\mbox{MeV}$ to $2.4\,\mbox{GeV}$
the suppression is {\it exclusively} due to the behavior of the three-gluon vertex.

\section{\label{sec:concl} Discussion and Conclusions}

In this article we have considered several nonperturbative aspects related to 
the gluon propagator, $\Delta(q^2)$, and the three-gluon vertex, $\Gamma_{\alpha\mu\nu}$, 
in the context of Landau gauge QCD with $N_f$=2+1 dynamical quarks. Our approach combines
a SDE-based analysis, carried out within the PT-BFM
framework, with new data gathered from lattice QCD simulations with $N_f$=2+1 domain wall fermions.
In particular, from the SDE point of view, the gluon kinetic term $J(q^2)$
has been computed indirectly, by obtaining $m^2(q^2)$
from its own ``gap equation'' and then ``subtracting'' it from the new lattice
data for $\Delta^{-1}(q^2)$. 
The  $J(q^2)$ so determined  is subsequently used
for the ``gauge technique'' reconstruction (BC solution) of certain 
key form factors of $\Gamma_{\alpha\mu\nu}$, evaluated
at two special kinematic configurations (``symmetric'' and ``asymmetric'').  
The two main quantities emerging from this construction,
denoted by $\overline{\Gamma}_1^{\rm  sym}(s^2)$   and  $\overline{\Gamma}_3^{\rm asym}(q^2)$,
are then compared with recently acquired lattice data,  
displaying very good coincidence. We emphasize that, while the determination of $J(q^2)$
hinges on the use of the lattice data for the gluon propagator,
the subsequent results derived by means of this $J(q^2)$
constitute genuine theoretical predictions. 

There are certain key theoretical notions underlying this work which are worth highlighting.

({\it i}) The recent nonlinear SDE analysis of~\cite{Aguilar:2019kxz}
generalizes from pure Yang-Mills to the case of real-world QCD 
with dynamical quarks, giving rise to a $m^2(q^2)$ that 
displays all qualitative features known from the quenched case. 

({\it ii}) The low-momentum behavior of $J(q^2)$ is clearly dominated by the
unprotected logarithm originating from the ghost loop.
In a pure Yang-Mills context, the diverging contribution of this logarithm
overcomes the opposing action of its protected counterparts,
leading to the IR suppression of $J(q^2)$ and its zero crossing. 
The inclusion of quark loops, which are regulated by the quark masses, 
gives rise to additional IR finite contributions, whose  
net effect is to attenuate the aforementioned outstanding features.

({\it iii})
By virtue of the fundamental STI of \1eq{stig}, the longitudinal form factors of $\Gamma_{\alpha\mu\nu}$ display the
same qualitative characteristics as the $J(q^2)$; in that sense,
the influence of the ghost sector, and in particular of the ghost-gluon kernel,
is rather limited, and does not alter the main dynamical properties that 
$\overline{\Gamma}_1^{\rm  sym}(s^2)$ and  $\overline{\Gamma}_3^{\rm asym}(q^2)$ inherit from the $J(q^2)$. 

({\it iv})
In our opinion, the present analysis provides additional support for the
picture of the IR sector of (Landau gauge) QCD that has emerged in recent years,
according to which, quarks and gluons acquire dynamically generated masses, while the   
ghosts remain strictly massless~\cite{Aguilar:2008xm,Boucaud:2008ky,Fischer:2008uz,Cloet:2013jya,Binosi:2014aea}. The three-gluon vertex appears to be the host of an  
elaborate synergy between the mechanisms responsible for this exceptional mass pattern,
thus providing an outstanding testing ground both for physics ideas as well as computational methods.

\acknowledgments 
We are very grateful to Ph. Boucaud for his crucial help in obtaining the numerical data, during the early stages of this project, and to the RBC/UKQCD collaboration, especially P. Boyle, N. Christ, Z. Dong, C. Jung, N. Garron, B. Mawhinney and O. Witzel, for access to the lattice configurations employed herein. Our lattice calculations benefited from the following resources: CINES, GENCI, IDRIS (Project ID 52271); and the IN2P3 Computing Facility. 
This work is supported by the Spanish Ministry of Economy and Competitiveness (MINECO) under grants FPA2017-84543-P (J.~P.) and FPA2017-86380-P (J.~R.-Q. and F.~S.).  J.~P. also acknowledges the Generalitat Valenciana for the  grant  Prometeo/2019/087.
The work of  A.~C.~A. and M.~N.~F. is supported by the Brazilian National Council for Scientific and Technological Development (CNPq) under the grants 305815/2015, 142226/2016-5, and 464898/2014-5 (INCT-FNA). A.~C.~A. also acknowledges the financial support
from  S\~{a}o Paulo Research Foundation (FAPESP) through the project 
2017/05685-2.

\end{document}